\documentclass[prl, aps, nofootinbib, twocolumn, amssymb, superscriptaddress, 10pt]{revtex4-2}
\usepackage{amsmath}
\usepackage{amssymb}
\usepackage{amsthm}
\usepackage{amsfonts}
\usepackage{enumerate}
\usepackage{latexsym}
\usepackage{color}
\usepackage{setspace} 
\usepackage{blindtext}
\usepackage{dsfont}
\usepackage{mathrsfs}
\usepackage{mathptmx} 
\usepackage[normalem]{ulem}

\usepackage{tabularx}
\newcolumntype{Y}{>{\centering\arraybackslash}X}

\usepackage{stackengine}

\usepackage{bm}
\usepackage{graphicx}

\usepackage{hyperref}
\hypersetup{
pdfnewwindow=true, colorlinks=true,
linkcolor=blue, anchorcolor=blue,
citecolor=blue, filecolor=blue,
menucolor=blue, urlcolor=blue} 

\usepackage{pifont}

\newcommand{\yf}[1]{{\color{blue} #1}}

\usepackage{verbatim}

\begin{document}

\author{Yuan Fang}
\affiliation{Department of Physics \& Astronomy,  Extreme Quantum Materials Alliance, Smalley-Curl Institute, Rice University, Houston, Texas 77005, USA}

\author{Lei Chen}
\affiliation{Department of Physics \& Astronomy,  Extreme Quantum Materials Alliance, Smalley-Curl Institute, Rice University, Houston, Texas 77005, USA}

\author{Andrey Prokofiev}
\affiliation{Institute of Solid State Physics, TU Wien, Wiedner Hauptstr.\ 8-10, 1040 Vienna, Austria}

\author{I\~{n}igo Robredo}
\affiliation{Donostia International  Physics  Center,  P. Manuel  de Lardizabal 4,  20018 Donostia-San Sebastian,  Spain}
\affiliation{Max Planck Institute for Chemical Physics of Solids, Noethnitzer Str. 40, 01187 Dresden, Germany}
\affiliation{Materials Research and Technology Department, Luxembourg Institute of Science and Technology (LIST), Avenue des Hauts-Fourneaux 5, L-4362 Esch/Alzette, Luxembourg}

\author{Jennifer Cano}
\affiliation{Department of Physics and Astronomy, Stony Brook University, Stony Brook, NY 11794, USA}
\affiliation{Center for Computational Quantum Physics, Flatiron Institute, New York, NY 10010, USA}

\author{Maia\ G.\ Vergniory}
\affiliation{Donostia International  Physics  Center,  P. Manuel  de Lardizabal 4,  20018 Donostia-San Sebastian,  Spain}
\affiliation{Département de physique et Institut quantique, Université de Sherbrooke, Sherbrooke, Québec, Canada J1K 2R1}

\author{Silke Paschen}
\affiliation{Institute of Solid State Physics, Vienna University of Technology, Wiedner 
Hauptstr. 8-10, 1040 Vienna, Austria}

\author{Qimiao Si}
\affiliation{Department of Physics \& Astronomy,  Extreme Quantum Materials Alliance, Smalley-Curl Institute, Rice University, Houston, Texas 77005, USA}

\title{Magnetic Weyl-Kondo semimetals induced by quantum fluctuations}

\begin{abstract}
Weyl-Kondo semimetals are strongly correlated topological semimetals that develop through the cooperation of the Kondo effect with space group symmetries. The Kondo effect, capturing quantum fluctuations associated with strong correlations, is usually suppressed by magnetic order. Here we develop the theory of magnetic Weyl-Kondo semimetal. The key of the proposed mechanism is that the magnetic order comes from conduction $d$ electrons, such that the local $f$ moments can still fluctuate. We illustrate the extreme case where the magnetic space group symmetries prevent any spontaneous magnetization on the sites with the $f$-orbitals. In this case, topological degeneracies, including hourglass Weyl-Kondo nodal lines, appear when the magnetic space group symmetry constrains the Kondo-driven low-energy excitations; they lead to a third-order nonlinear anomalous Hall response. Based on the proposed mechanism, we explore the interplay between strong correlations and symmetries with database search leading to several candidate materials. The most prominent candidates are antiferromagnetic $\rm UNiGa$ and $\rm UNiAl$, with a third-order anomalous Hall response, as well as ferromagnetic $\rm USbTe$ and $\rm CeCoPO$, with a first-order one. Our findings pave the way for future experimental and theoretical investigations that promise to further advance the overarching theme of strongly correlated topology.
\end{abstract}

\maketitle

\emph{Introduction.}---
Strong correlations drive a variety of quantum phases~\cite{Kei17.1,Pas21.1}. 
Weyl-Kondo semimetals (WKSMs)~\cite{lai2018weyl,dzsaber2017kondo} represent a fascinating intersection of the strong correlation physics with gapless electronic topology~\cite{Armitage2017}.
This interplay leads to a rich platform of emergent phenomena and exotic physical properties.

Weyl semimetals in noninteracting systems have emerged as an important class of topological materials characterized by the topology of Weyl nodes, either in paramagnetic~\cite{Armitage2017,son2013chiral,hosur2013recent,zyuzin2012topological} or magnetic environments~\cite{yan2017topological,elcoro2021magnetic}.
These nodes act as sources or sinks of Berry curvature, giving rise to intriguing phenomena such as chiral anomaly and Fermi arc surface states. In realizing Weyl semimetals, 
lattice symmetries play an important role as indicators of topology~\cite{Armitage2017,cano2021band}. On the other hand, Kondo physics describes the screening of localized magnetic moments by conduction electrons, leading to the formation of composite heavy fermions~\cite{Kirchner2020,Hewson1997}.

In WKSMs, these two paradigms coalesce to produce a novel class of materials with distinct physical properties. The cooperation between strong electron correlations and space group symmetry constraints can induce the Weyl nodes in the heavy fermion states, as revealed by recent non-perturbative studies~\cite{chen2022topological,lai2018weyl,grefe2020weyl,grefe2020extreme,Hu-Si2021}.
Such novel phases have been explored in experiments in the material $\rm Ce_3Bi_4Pd_3$, based on measurements of specific heat and spontaneous Hall effect~\cite{dzsaber2017kondo,dzsaber2021giant,dzsaber2022control}.
Potential materials candidates of WKSMs have been proposed thorough database-mining based on both space group symmetries and the reported physical properties~\cite{chen2022topological}.

WKSMs have so far been confined to paramagnetic settings. Yet, heavy fermion systems are often on the verge of magnetic order; indeed, their interplay with magnetism is a prominent feature of the field. 
We are thus motivated to study magnetic Weyl-Kondo semimetals.
To make progress, it is necessary to understand how the Kondo effect can operate in a magnetic environment given that the latter usually inhibit the local moments' capacity to produce quantum fluctuations.

In this work, we confront this challenge and develop the theory of magnetic Weyl-Kondo semimetal (MWKSM). While a magnetic order breaks time-reversal symmetry, a suitable combination of the time-reversal symmetry and spatial symmetries can be preserved. Such symmetries are described by the magnetic space groups (MSGs), which have been classified and tabulated~\cite{bradley2010mathematical}. 
Here we demonstrate the development of MWKSM from the quantum fluctuations of the local moments, with a particular emphasis on the case when the magnetic order is driven by the conduction $d$ electrons.
Our model calculations give rise to Kondo-driven magnetic 
Weyl semimetals, including the hourglass Weyl nodal lines (HWNLs).
To be definite, we first  focus on the MWKSMs in the square-net category given the large materials base in this category for topological semimetals~\cite{CanoSquareNet,klemenz2019topological,klemenz2020role,lei2023weyl}. This will be followed by searching for generic $\rm Ce$, $\rm U$ or $\rm Yb$ based materials.
Since Weyl points and HWNLs are sources of Berry curvature, we find that the nonlinear anomalous Hall effect~\cite{zhang2023higher,fang2023quantum,sodemann2015quantum} provides telltale signatures of the symmetries that are broken in the MSGs for the MWKSMs.

Specifically, we study the Anderson/Kondo lattice model in several square-net MSGs, with a particular focus on MSG no.~129.415 ($P4'/nmm'$) to illustrate our case. (We use Belov-Neronova-Smirnova (BNS) setting~\cite{bradley2010mathematical}.)
In the presence of the $d$-electron-driven magnetic order, the $f$-electron local moment can still fluctuate because the space group symmetry prevents any spontaneous magnetization on the sites with the $f$-orbitals.
In turn, we show how, through the space-group symmetry constraints~\cite{PhysRevMaterials.3.054203}, Kondo-driven  HWNLs develop and induce a 3rd-order nonlinear Hall response.
We also report related results for several other MSGs.
Armed with the theoretical results, we outline a materials design procedure for MWKSMs with symmetry-constrained topological crossings, and identify a set of candidate materials with strong correlations that are expected to feature the symmetry enforced non-linear anomalous Hall responses.
Of particular interest are antiferromagnetic (AFM) $\rm UNiGa$ and $\rm UNiAl$, which are predicted to show third-order anomalous Hall responses, as well as ferromagnetic (FM) $\rm USbTe$ and $\rm CeCoPO$, for which first-order anomalous Hall responses are expected.

\emph{MWKSM in a magnetic square net model}---
To realize the Kondo effect, we assume there are $\rm Ce$ (or $\rm U$, $\rm Yb$, containing $f$ electrons) and transition metal ($\rm TM$) atoms.

\begin{figure}[t!]
    \centering
\includegraphics[width=0.7\linewidth]{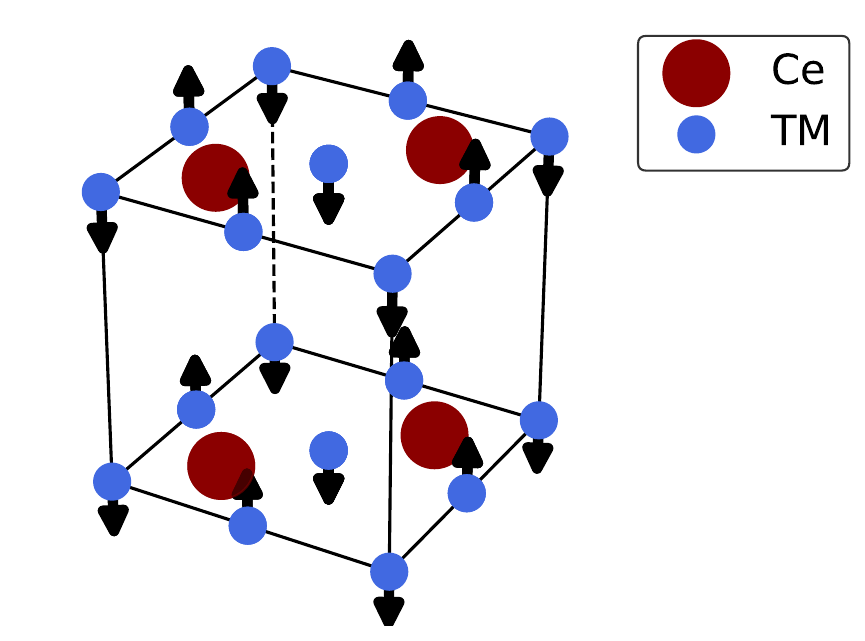}
    \caption{Unit cell of a model material in MSG no.~129.415 ($P4'/nmm'$). The AFM ordering on the $\rm TM$ atoms are illustrated by arrows. The correlated $f$-electrons come from the $\rm Ce$ atoms.}
    \label{fig:unit_cell}
\end{figure}

We provide a proof-of-principle demonstration of this MWKSM setup in the square-net case with MSG  no.~129.415 ($P4'/nmm'$).
The $\rm Ce$ atoms are at the $2a$ Wyckoff position and the $\rm TM$ atoms are at the $4d$ Wyckoff position. The magnetic moments on the $\rm TM$ atoms are ordered, as shown in Fig.~\ref{fig:unit_cell}, due to interactions among the $d$ electrons.
The $C_4{\cal T}$ symmetry ensures that no magnetic ordering can appear at the $2a$ Wyckoff position, which forbids the $f$ electrons of the $\rm Ce$ atoms from developing a spontaneous magnetic moment.

The conduction electrons are, as standard, taken as non-interacting, with the following Hamiltonian:
\begin{align}
\label{eqn:Hc}
    H_c =& \sum_k \Psi_k^\dagger \left( h({\mathbf k}) - \mu \right) \Psi_k \\
    h({\mathbf k}) &= t \cos(\frac{k_x}{2})\cos(\frac{k_y}{2}) \sigma_0\tau_x + t_z \sin k_z \cos(\frac{k_x}{2})\cos(\frac{k_y}{2}) \sigma_0\tau_y \nonumber\\
    &+t_{SOC} \big( \sin k_x\sigma_y+\sin k_y\sigma_x \big) \tau_z  \nonumber\\
    &+t_z' \sin k_z \left( \cos k_x - \cos k_y \right) \sigma_0 \tau_z
    +t_{SOC}' \sin(\frac{k_x}{2})\sin(\frac{k_y}{2}) \sigma_z\tau_x  \nonumber\\
    &+t_{SOC}'' \left( \cos(\frac{k_x}{2})\sin(\frac{k_y}{2}) \sigma_x - \sin(\frac{k_x}{2})\cos(\frac{k_y}{2}) \sigma_y \right) \tau_y \, ,
\label{eqn:129.415}
\end{align}
where $\mu$ is chemical potential, $\Psi = (c_{A\uparrow},c_{A\downarrow},c_{B\uparrow},c_{B\downarrow})$ is the annihilation operator of itinerant electrons at the two $\rm Ce$ sublattice sites $A$ and $B$ with the two spin degrees, $\sigma$ and $\tau$ are the Pauli matrices that describe the spin and sublattice degrees respectively.
The first and second lines of $h({\mathbf k})$ preserve $P4/nmm1'$. Meanwhile, the third and fourth lines preserve $P4'/nmm'$, which are responsible for the spin splitting; these terms can be considered as a form of spin-orbit coupling (SOC) enabled by the magnetic environment that the electrons from the $\rm TM$ atoms provide.

The $f$ electrons are at the energy level $\epsilon_f$ and have an on-site Hubbard interaction $U$.
In order to study the symmetry-enabled Kondo screened phase, we consider the soluble case realized in the large $U$ limit; here, the model is solved at the saddle-point level, which formally develops in a large-$N$ limit and in terms of a slave-boson representation~\cite{Hewson1997}. The effective Hamiltonian is given as follows:
\begin{equation}
\label{eqn:model}
\begin{split}
    H=H_c+\sum_{j,\alpha,m} \epsilon_f f_{j\alpha m}^\dagger f_{j\alpha m} +rV \left[c_{j\alpha m}^{\dagger} f_{j\alpha m}+\text { h.c. }\right] \\
    + \lambda \sum_j (n_{f,j}+r^2-Q)    \, . 
\end{split}
\end{equation}
Here, $r=\langle b_j \rangle$ is the condensed slave boson field $b_j$, while $c_{j\alpha m}$ and $f_{j\alpha m}$ represent the conduction and heavy electrons at site $j$, orbital $\alpha$, spin $m$ while $n_{f,j}=\sum_{\alpha m}f^\dagger_{j\alpha m}f_{j\alpha m}$ represents the number of $f$ electrons at site $j$.
The filling is constrained to $Q=1$.  
The $\lambda$ term comes from the Lagrange multiplier of the local filling constraint $n_{f,j}+b^\dagger b=Q$.
The self-consistent equations read as follows:
\begin{align}
    \langle n_{f,j}\rangle + r^2 &= Q \label{eqn:self1}\\
    V \sum_{\alpha,m}\langle c_{j\alpha m}^{\dagger} f_{j\alpha m}\rangle  &= -2\lambda r \label{eqn:self2}
\end{align}
where $\langle\cdot\rangle$ means averaging over all sites.
We solve the self-consistent equations 
by iteration to find the physical values of $\lambda$ and $r$. During this calculation, we set $\mu = -r^2V^2/\lambda$ to maintain the physical filling factor.
Without loss of generality, we consider the case with the parameters $t=1$ (defined as the energy unit), $t_z=0.5$, $t_{SOC}=0.5$, $t_{SOC}'=2$, $t_{SOC}''=1$, $t_z'=0.5$, $V=10$, $\epsilon_d=-5$.
The self-consistent calculation yields $r^2=0.314$, $\lambda=2.946$ and $\mu=-10.6$. 

\begin{figure} [t!]
    \centering
    \includegraphics[width=\linewidth]{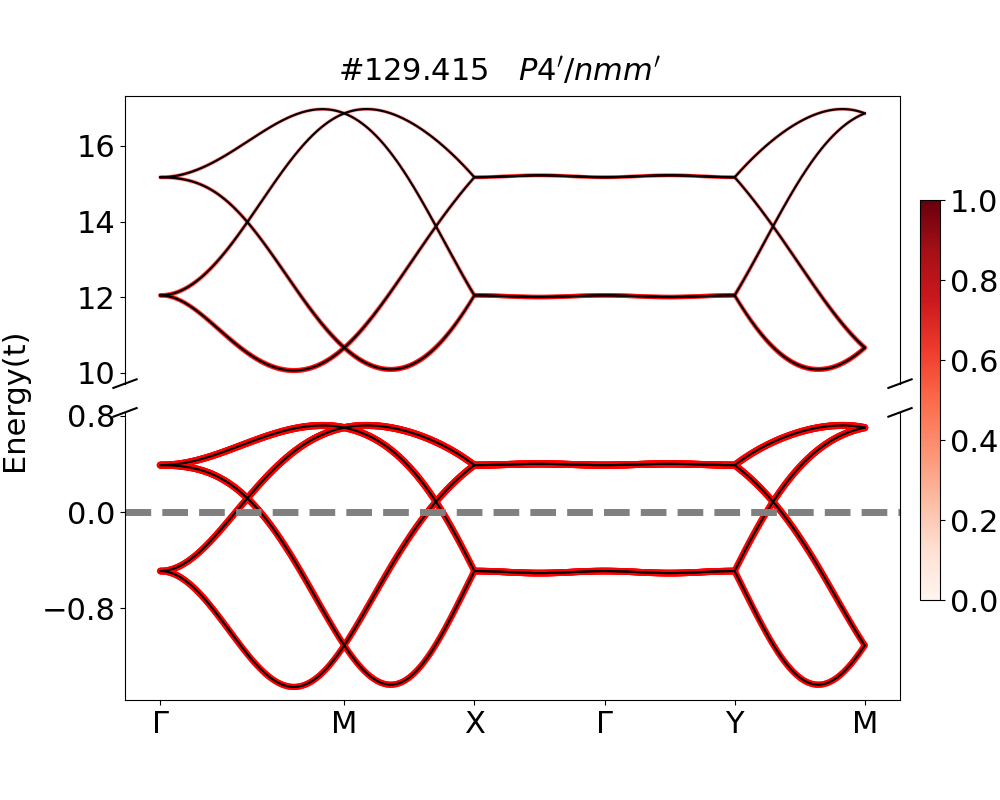}
    \caption{Kondo-driven Weyl hourglass nodal line excitations.
   Energy vs. the wavevector $\bf{\rm k}$ is shown along a high-symmetry direction. Red color indicates the weight of the $f$ electrons in the state. The crossing extends to a nodal line structure as shown in Fig.~\ref{fig:FS_BC} (a).}
    
    \label{fig:129.415}
\end{figure}

Fig.~\ref{fig:129.415} shows the energy dispersion at $k_z=0$. The top four dispersive wide bands mainly describe the conduction electrons, with symmetry-enforced crossings that occur far away from the Fermi energy, as it generically should be.
On the other hand, the bottom four narrow bands are Kondo-driven, capturing the composite heavy fermions that develop from the Kondo effect. 
Because these states are emergent, driven by the strong interactions, they lie at low energies and, as such, they are bound to the Fermi energy.

\begin{figure}[t!]
    \centering
    \includegraphics[width=\linewidth]{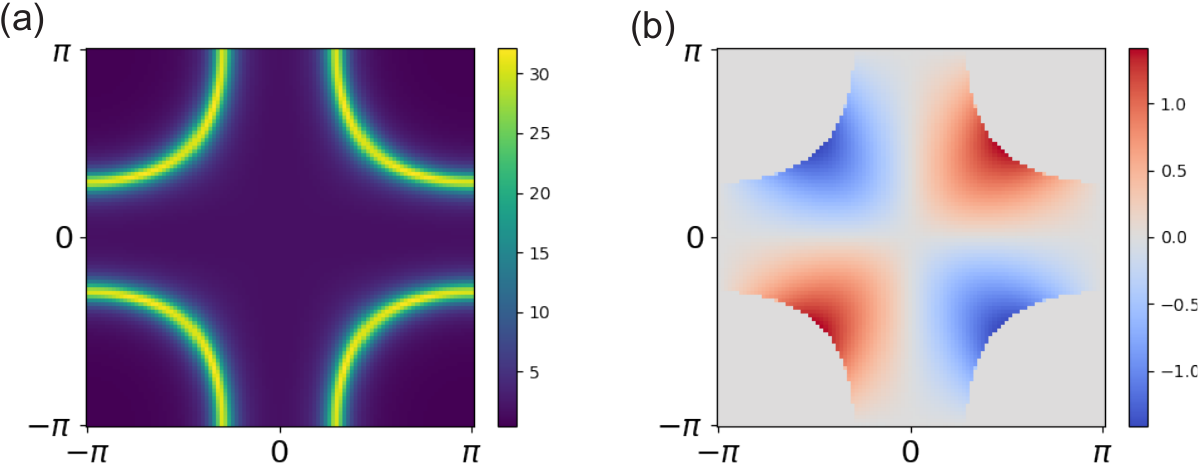}
    \caption{(a) Spectral function $A(k,\omega=0.1)$ shows the Kondo-driven HWNLs of the 
    $f$-electron states of 
    Fig.~\ref{fig:129.415}, where 
    the HWNLs reside. (b) Berry curvature of the 
    states below the Fermi energy, showing 
    a quadrupole-shaped distribution.}
    \label{fig:FS_BC}
\end{figure}

These emergent states are subject to symmetry constraints. Accordingly, hourglass Weyl Kondo nodal lines form, which are shown in Fig.~\ref{fig:FS_BC} along with the Berry curvature of states below the Fermi energy. The Berry curvature distribution has hot spot around these nodal lines. This distribution is quadrupole-shaped.
A 3rd-order anomalous Hall response is driven by Berry curvature quadrupoles, and is defined by $j^\mu = \sigma^{\mu;\alpha\beta\gamma} E_\alpha E_\beta E_\gamma$. It has $1\omega$ and $3\omega$ responses where $\omega$ is the frequency of the electric field. 
The 3rd-order response conductivity captures the second order derivative of the Berry curvature, i.e. Berry curvature quadrupole, which leads to an in-plane current $j^x \propto E_xE_y^2$~\cite{zhang2023higher}.

The Berry phases, analyzed through the Wilson loops, imply the existence of mid gap surface states at the Fermi energy that are localized on the boundaries of finite systems. These are discussed in the SM~\cite{sm}.

The HWNLs result from compatibility conditions in momentum space. For example, in MSG no.~129.415 ($P4'/nmm'$) the compatibility condition forces the irreducible representations (irreps) connectivity: $\overline{\Gamma}_5(2)\rightarrow \overline{D}_3(1)\oplus \overline{D}_4(1)$, $\overline{\Gamma}_6(2)\rightarrow \overline{D}_3(1)\oplus \overline{D}_4(1)$ and $\overline{M}_3(2)\rightarrow \overline{D}_4(1)\oplus \overline{D}_4(1)$, $\overline{M}_4(2)\rightarrow \overline{D}_3(1)\oplus \overline{D}_3(1)$, where $D=(u,v,0)$, $\Gamma=(0,0,0)$, $M=(\pi,\pi,0)$ and the number in parenthesis indicates the dimensionality of the irrep.
They imply a crossing along any
path on the $k_z=0$ plane that connects $\Gamma$ and $M$, i.e. there are symmetry enforced HWNLs in the $k_z=0$ plane~\cite{bradley2010mathematical,elcoro2021magnetic,xu2020high}. 

This illustrating example shows that MWKSMs can have symmetry enforced crossings and nonlinear Hall effects. In the SM~\cite{sm} we show more examples of MSGs in the space group no.~129 family. With these solutions in place, we turn to a general discussion about how to identify these physical properties from symmetry considerations.

\emph{Magnetic space groups and physical properties.}\label{sec:MSG}---
Our calculations described above illustrate an important point for magnetic heavy fermion systems.
Provided the Kondo effect is able to operate, the emergent composite heavy fermions are bound to the Fermi energy even in a magnetic environment. We will thus utilize symmetry considerations to search for suitable materials and realize the MWKSMs.
Of our interest are a number of physical properties, which are all constrained by symmetries. These include 
(1) the symmetry enforced crossings; 
(2) symmetry constraints on the magnetic ordering;
and 
(3) symmetry constraints on the nonlinear Hall response.
These are described in some detail in the SM~\cite{sm}.

An example is the 129 family of MSGs in Table~\ref{tab:tab129}. 
Specifically, MSG no.~129.417 is compatible with FM ordering while all MSGs in this family, except no.~129.412, are compatible with certain AFM ordering. MSGs no.~129.411, 129.414, 129.415 have 3rd-order Hall response as the leading order.
\begin{table}[ht]
    \centering
    \begin{tabular}{cc|c|c|c |c}
         MSG \#&BNS  & FM & AFM & Hall response &Crossing\\
         \hline 
         129.411 &$P4/nmm$ & &\checkmark & 3rd-order & Dirac\\
         129.412 &$P4/nmm1'$ & & &  & Dirac\\
         129.413 &$P4/n'mm$ & &\checkmark &  & Dirac\\
         129.414 &$P4'/nm'm$ & &\checkmark & 3rd-order & Dirac\\
         129.415 &$P4'/nmm'$ & &\checkmark & 3rd-order & HWNL\\
         129.416 &$P4'/n'm'm$ & &\checkmark & &  \\
         129.417 &$P4/nm'm'$ & \checkmark &\checkmark & 1st-order  &  \\
         129.418 &$P4'/n'mm'$ & &\checkmark & & Dirac \\
         129.419 &$P4/n'm'm'$ & &\checkmark & &  \\
         \hline
         129.420 &$P_{c}4/nmm$ & &\checkmark & & Dirac\\
         129.421 &$P_{C}4/nmm$ & &\checkmark &  & Dirac\\
         129.422 &$P_{I}4/nmm$ & &\checkmark &  & Dirac\\
    \end{tabular}
    \caption{MSGs of space group no.~129 family. We use BNS setting to denote the MSGs~\cite{bradley2010mathematical}. 
    The magnetic orderings compatible with each MSG are marked with $\checkmark$. Blank means the magnetic ordering is not compatible with the group. The leading order Hall response of each group is listed. Blank means no Hall response of any order is permitted. The crossing types for bands from $s$-orbitals at $2a$ Wyckoff positions are listed. Blank means no symmetry-enforced Dirac or HWNL exists in the group.
    }
    \label{tab:tab129}
\end{table}
In Table.~\ref{tab:tab129} we also list the crossings other than Weyl points and gapped phases. 
Weyl points can appear in all of these groups except MSG no.~129.412, 129.413, 129.416 and 129.418-129.422 where ${\cal PT}T_{\mathbf r}$ symmetry is present ($\cal P$ is inversion, $\cal T$ is time-reversal symmetry, and $T_{\mathbf r}$ indicates a fractional translation).
Symmetry enforced Dirac points and HWNLs can appear in some of these groups. As is also seen, the Hall responses of orders one and three as leading order can be found for different MSGs. We also list MSG families 125 and 31 for comparison in the SM~\cite{sm}.

We emphasize that the 3rd order anomalous Hall response of the MWKSMs represents a novel observable.
First, the magnetic ordering that breaks both time-reversal and inversion symmetries allows the 3rd-order response.
Second, Weyl nodal lines are the sources of Berry curvature multipoles. Since the Kondo effect pins the Weyl nodal lines close to the Fermi energy, the response will be strong.

\emph{Materials candidates}\label{sec:candidates}---
To set the stage for the identification of candidate materials, we search for Ce, U and Yb based compounds, with special focus on square net cases~\cite{klemenz2019topological,klemenz2020role,CanoSquareNet}, which are conducive to Weyl-Kondo semimetals~\cite{lei2023weyl}. We employ a systematic approach from a database search with the following strategy:
\\
(i) Magnetic ordering:
The magnetic ordering is preferable if they are coming from $d$ electrons, that has a sufficiently high Curie temperature or N{\'e}el temperature.
To broaden the search, we will also consider a similar setting, as exemplified by $U$-based systems, in which each $U$ ion can contribute multiple $f$ electrons; orbital-selective correlations can make some $f$-orbitals to produce magnetic order, while the remaining $f$-orbital can still develop the Kondo effect.
We are in particular interested in the MSGs that have nonlinear Hall effects as the leading order responses. 
\\
(ii) Strong electron correlation:
The Kondo physics captures the quantum fluctuations associated with the underlying strong correlations. Where possible, we filter the materials based on specific heat measurements. A large ``coarse-filtered" 
Sommerfeld coefficient $\gamma$, determined from $C_p/T$ at low temperatures (where $C_p$ is the specific heat), serves as a proxy for enhanced quantum fluctuations driven by electronic correlations. 
\\
(iii) Semimetallicity:
Materials with semimetallic behaviors at low temperature, as evidenced by a large resistivity or a small carrier concentration, are potential Kondo semimetals.

We summarize the screened materials in detail in the SM~\cite{sm} and list the primary candidate materials below based on the order of the anomalous Hall effect.

The candidates with expected 3rd order response include $\rm UNiGa$ and $\rm UNiAl$~\cite{UNiGa,UNiGamag}. These two materials have AFM order with MSG no~189.224 ($P\bar 6'2m'$). 
The temperature dependence of the resistivity indicates semimetallic behaviors (see SM~\cite{sm}). These two materials are expected to have symmetry enforced Weyl points.

The candidates with 1st order response include $\rm CeCoPO$~\cite{CeCoPO} and $\rm USbTe$~\cite{USbTe},
which feature magnetic orderings driven by $d$- and orbital-selective $f$-electrons, respectively. 

$\rm CeCoPO$ has FM order and is thus expected to show linear anomalous Hall effect. 
The carrier concentration is around $1\times10^{21} {\rm cm}^{-3}$, as estimated from measurements in the similarly-behaving material $\rm CeFeAsO$~\cite{Yuan_2011,MagneticStudy_CeFeAsOF,Jaroszynski2008, review2016Si}; measurements of the Hall and optical conductivity 
in $\rm CeCoPO$ will help further establish its semimetal nature.
Its MSG is no.~129.417 ($P4/mm'm'$).
There are symmetry enforced degeneracy at Brillouin zone boundary and potentially symmetry protected Weyl nodal lines at $k_z=0,\pi$ planes. 

In order to identify the nodal features, we performed density functional theory (DFT) calculations as implemented in full-potential local-orbital (FPLO) code \cite{FPLO1,FPLO2}. We then constructed maximally localized Wannier functions to study the Fermi-surface cuts at the $k_z=0$ plane, verifying the symmetry-constrained Weyl nodal lines~\cite{sm}.
The degeneracies and nodal lines that happen away from Fermi energy are consistent with our symmetry analysis. Similar symmetry considerations apply to the $f$-electron states close to Fermi energy which we show in the toy model calculations~\cite{sm}.

$\rm USbTe$~\cite{USbTe} has FM order and is thus expected to possess linear anomalous Hall effect. Its Curie temperature is 125K.
The Sommerfeld coefficient is enhanced, especially in light of its low carrier concentration $6\times 10^{20}~{\rm cm}^{-3}$~\cite{USbTe}). The resistivity of $\rm USbTe$ shows semimetallic behavior (see SM~\cite{sm}). Its MSG is no.~129.417 ($P4/mm'm'$). There are symmetry enforced degeneracy at Brillouin zone boundary and potentially symmetry protected Weyl nodal lines at $k_z=0,\pi$ planes. 

The $f$ electrons in the four materials are not immune to magnetic orderings in their MSGs. However, since the magnetic moment is weak, the heavy fermion phase still survives.

In addition to these promising candidates, we found several interesting candidates. Either the semimetallicity or the Sommerfeld coefficient is unknown. Therefore, we list the information in SM~\cite{sm}. Further investigation of the physical properties of these materials will be illuminating.

\emph{Discussion.}---
We remark on several important points. Firstly, in order to perform calculations for a substantial number of MSGs, we have treated the Anderson lattice models at a saddle-point level. Going beyond this level, the quasiparticles remain robust but will acquire damping. The topological crossings we have shown will be robust based on the symmetry constraints of Green's function eigenvectors~\cite{Hu-Si2021}.

Secondly, we have emphasized the signatures in terms of the nonlinear anomalous Hall effect, because this quantity has played a key role in the identification of the WKSMs in the paramagnetic settings, as studied experimentally~\cite{dzsaber2021giant} and theoretically~\cite{grefe2020weyl}.
Still, other experimental signatures can be considered. For example, the hourglass Weyl Kondo nodal lines are expected to manifest both in the temperature dependence of the electronic specific heat~\cite{lai2018weyl,dzsaber2017kondo} and in the STM spectrum~\cite{chen2022topological}.

Finally, our materials search has identified a number of Ce- and U-based square-net heavy-fermion materials in which the electronic topology is manifested in either first-order or third-order nonlinear anomalous Hall effects. 
The latter has never been considered before for heavy fermion systems or, to our knowledge, any other magnetic metals showing strong correlations. We hope that our proposal will motivate such experiments in these broad classes of quantum materials.

To summarize, we have developed the theory of magnetic Weyl-Kondo semimetal. We formulate the theory in terms of a magnetic order that originates from the conduction ($d$) electrons, such that the $f$-electron local moments are still able to fluctuate and strong correlations can still induce amplified quantum fluctuations in spite of the static magnetic order.
We have illustrated how the symmetries of the magnetic space groups play an essential role in several regards. These include enabling the $f$-electron quantum fluctuations in magnetic settings, constraining topological crossings of the Kondo-driven low-energy excitations, as well as giving rise to nonlinear anomalous Hall effects.
Our results allow us to advance a design procedure for the magnetic Weyl-Kondo semimetals. The candidate materials we have proposed allow for experimental measurements that have never been attempted before in magnetic metals with extreme strong correlations.
As such, our work not only advances the enormously challenging field of strongly correlated topological matter but also opens new directions for the investigations on the venerable interplay between strong quantum fluctuations and electronic orders through the new lenses of correlated gapless topology.

\emph{Acknowledgement.}---
Work at Rice has primarily been supported by the Air Force Office of Scientific Research under Grant No.
FA9550-21-1-0356 (conceptualization and model construction, Y.F.),
by the National Science Foundation
under Grant No. DMR-2220603 (model calculations, Y.F. and L.C.),
by the Robert A. Welch Foundation Grant No. C-1411 and the Vannevar Bush Faculty Fellowship ONR-VB N00014-23-1-2870 (Q.S.). The
majority of the computational calculations have been performed on the Shared University Grid
at Rice funded by NSF under Grant EIA-0216467, a partnership between Rice University, Sun
Microsystems, and Sigma Solutions, Inc., the Big-Data Private-Cloud Research Cyberinfrastructure
MRI-award funded by NSF under Grant No. CNS-1338099, and the Extreme Science and
Engineering Discovery Environment (XSEDE) by NSF under Grant No. DMR170109. 
A.P. and S.P. acknowledge funding by the European Union (ERC, CorMeTop, project 101055088).
I.R. and M.G.V. acknowledge support to the  Spanish Ministerio de Ciencia e Innovacion (grant PID2022-142008NB-I00), partial support from European Research Council (ERC) grant agreement no. 101020833 and the European Union NextGenerationEU/PRTR-C17.I1, as well as by the IKUR Strategy under the collaboration agreement between Ikerbasque Foundation and DIPC on behalf of the Department of Education of the Basque Government and the Ministry for Digital Transformation and of Civil Service of the Spanish Government through the QUANTUM ENIA project call - Quantum Spain project, and by the European Union through the Recovery, Transformation and Resilience Plan - NextGenerationEU within the framework of the Digital Spain 2026 Agenda.
M.G.V. and S.P. acknowledge funding from the Deutsche Forschungsgemeinschaft (DFG, German Research Foundation) and the Austrian Science Fund (FWF) through the project FOR 5249 (QUAST).
M.G.V. received financial support from the Canada Excellence Research Chairs Program for Topological Quantum Materials.
J.C. acknowledges the support of
the National Science Foundation under Grant No. DMR-1942447, support from the Alfred P.
Sloan Foundation through a Sloan Research Fellowship and the support of the Flatiron Institute,
a division of the Simons Foundation. 
All authors acknowledge 
the hospitality of the Kavli Institute for Theoretical Physics, UCSB,
supported in part
by the National Science Foundation under Grant No. NSF PHY-1748958,
 during the program ``A Quantum Universe in
a Crystal: Symmetry and Topology across the Correlation Spectrum."  
Q.S. also 
acknowledges the hospitality of the Aspen Center for Physics, which is supported by the National Science Foundation under Grant No. PHY-2210452.

\bibliographystyle{apsrev4-2}
\bibliography{reference.bib}

\newpage \clearpage

\onecolumngrid
\setcounter{secnumdepth}{3}
\appendix
\begin{center}
	{\large
Magnetic Weyl-Kondo semimetals
induced by quantum fluctuations
	\vspace{4pt}
	\\
	SUPPLEMENTAL MATERIAL
	}
\end{center}

\section{\label{app:msg-symmetry} MSGs -- Symmetry considerations}

The symmetry groups compatible with magnetic orderings are described by MSGs. Due to the existence of magnetic moment, time-reversal symmetry ${\cal T}$ is broken. However, some of the products of spatial symmetries and ${\cal T}$ can be preserved. There are in total 1651 MSGs including 230 SGs without ${\cal T}$, 230 grey groups (SGs with ${\cal T}$), 674 type III MSGs (SGs with spatial-time-reversal symmetry except for translation-time-reversal symmetry) and 517 type IV MSGs (SGs with translation-time-reversal symmetry).

The magnetic ordering is determined by both the MSG and the site-symmetry group of the magnetic moment. The latter constrains the orientation of magnetic moment on each site while the former constrains the relation between the magnetic moments on different sites. Mathematically speaking, given a MSG $\cal G$ and a position $\mathbf q$ with its site symmetry group $G_{\mathbf q}$, we have the coset-decomposition ${\cal G} = \bigcup_{\alpha} g_\alpha G_{\mathbf q}$ where $g_\alpha$ can be either unitary or anti-unitary symmetries. For a compatible magnetic ordering, the magnetic momentum $\mathbf m$ of a certain orientation located at $\mathbf q$ is invariant under $G_{\mathbf q}$, forming an irreducible representation of $G_{\mathbf q}$. Then unitary $g_\alpha=\{R_{g_\alpha}|{\mathbf t}_{g_\alpha}\}$ maps 
\begin{equation}
    \mathbf q,~\mathbf m \mapsto R_{g_\alpha}\mathbf q + {\mathbf t}_{g_\alpha},~R(g_\alpha)\mathbf m~,
\end{equation}
and anti-unitary $g_\alpha=\{R_{g_\alpha}|{\mathbf t}_{g_\alpha}\}{\cal T}$ maps them to 
\begin{equation}
    \mathbf q,~\mathbf m \mapsto R_{g_\alpha}\mathbf q + {\mathbf t}_{g_\alpha},~-R(g_\alpha)\mathbf m~.
\end{equation}
Therefore, the magnetic ordering is uniquely determined by $\cal G$ and $G_{\mathbf q}$.

The nonlinear Hall responses result from Berry curvature multipoles. The $n$-th order nonlinear Hall response is $j^{\mu}=\sigma^{\mu;\alpha_1\dots\alpha_n}E_{\alpha_1}\dots E_{\alpha_n}$. This nonlinear Hall conductivity is determined by the $n$-th moment of Berry curvature 
\begin{equation}
    Q^{(n)}_{\beta_1\dots\beta_{n-1};\delta}=\sum_{n}\int_{\text{BZ}}f_{n}^{(0)}\partial_{\beta_1}\dots \partial_{\beta_n-1}\Omega_n^\delta~,
\end{equation}
where $f_n^{(0)}$ is the Fermi-Dirac distribution of the $n$-th band and $\Omega_n$ is the Berry curvature of that band.
The Berry curvature multipoles are determined by the magnetic point group of the MSG $\cal G$, which is the quotient group of MSG mod the lattice translation group, i.e. ${\cal G}/{\mathbb T}$~\cite{bradley2010mathematical}. Whether the Berry curvature multipoles of each orders vanish or not can be determined by checking the transformation of Berry curvature multipoles under the magnetic point group~\cite{fang2023quantum}. If under a symmetry $g\in{\cal G}$, $g Q^{(n)} \neq Q^{(n)}$, then the unequal components must vanish $Q^{(n)}_{\beta_1\dots\beta_{n-1};\delta} = 0$. This implies the following symmetry constraint on Berry curvature multipoles:
\begin{equation}
    Q^{(n)} = \frac{1}{|{\cal G }/{\mathbb T}|}\sum_{g\in {\cal G}/{\mathbb T}} g Q^{(n)}
\end{equation}
If all the components of order $n$ vanish, then there is no $n$-th order Hall response in this MSG. From this analysis, the leading order Hall responses of MSGs can be determined.

MSGs can be viewed as symmetry breaking of the grey groups, which are the space groups with time-reversal symmetry. The crystal potentials that break the grey group symmetries lead to different SOCs in the band structure according to the corresponding MSGs. 
The SOCs can create various crossing types especially in these nonsymmorphic groups.

In the grey group MSG no.~129.412, the symmetry enforced Dirac point is pinned at high symmetry momenta by the interplay between time-reversal, mirror and glide symmetries which only allow four-dimensional irreps. For other MSGs with symmetry enforced Dirac points, they are protected by the unbroken symmetries at those momenta. The detailed reasoning can be made by checking the Frobenius-Schur index of the irreps of the maximal unitary group of the little co-group~\cite{bradley2010mathematical,elcoro2021magnetic}.

\section{Details of the tight binding model in MSG no.~129.415}
\begin{figure}[t!]
    \centering
    \includegraphics[width=0.8\linewidth]{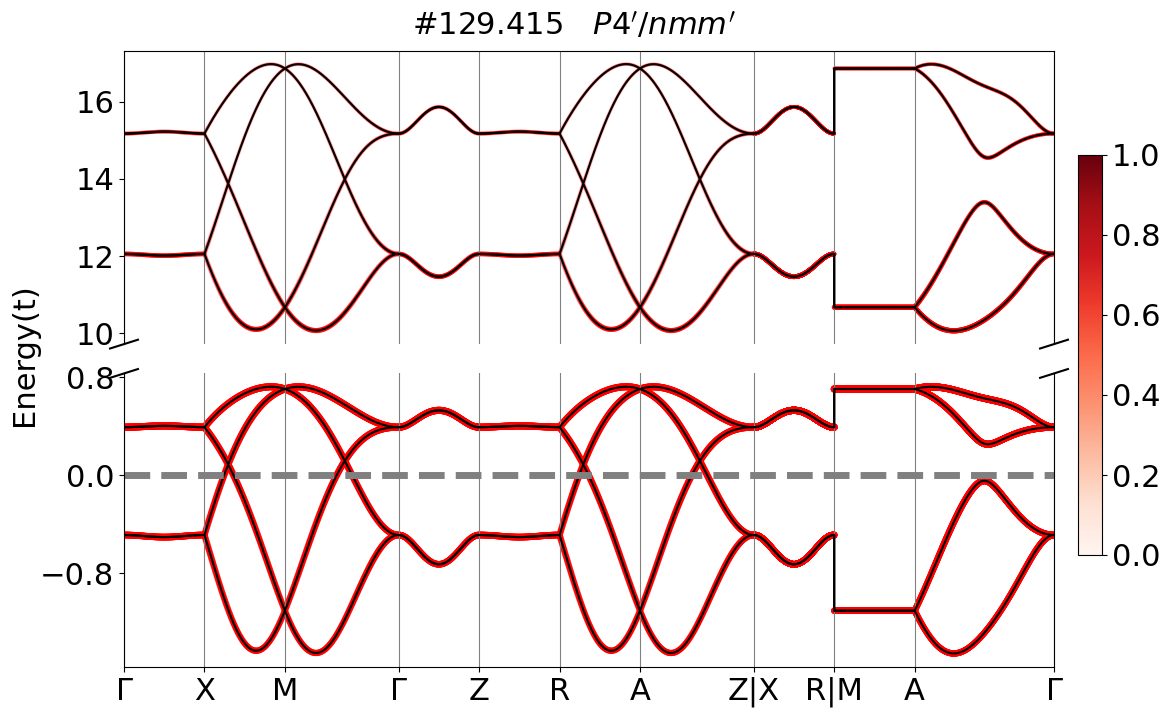}
    \caption{Energy dispersion of the Kondo-driven states along high symmetry lines in the three-dimensional Brillouin zone.}
    \label{fig:129_415_3D}
\end{figure}

The model we introduce in the main text is a three-dimensional model. In the main text, we
presented the band structure of the $k_z=0$ plane. The full Brillouin zone band structure is given here in Fig.~\ref{fig:129_415_3D}.
In this model, $k_z=0$ and $k_z=\pi$ 
have the same features: they both have hourglass Weyl nodal lines (HWNLs) surrounding the $(\pi,\pi)$ point; the Berry curvature distributions are both quadrupole shaped with the same signs. In the planes of $k_z\neq 0,\pi$, there are no HWNLs, as the bands along $A\Gamma$ line show. 
This reflects the mirror symmetry $m_{(001)}$ protecting the crossings at $k_z= 0,\pi$. 
There are two-fold degenerate bands along $\Gamma X$, $\Gamma Z$, $ZT$, $XR$, $MA$ due to the interplay between the mirror and glide symmetries.

\begin{figure}[ht]
    \centering
    \includegraphics[width=0.6\linewidth]{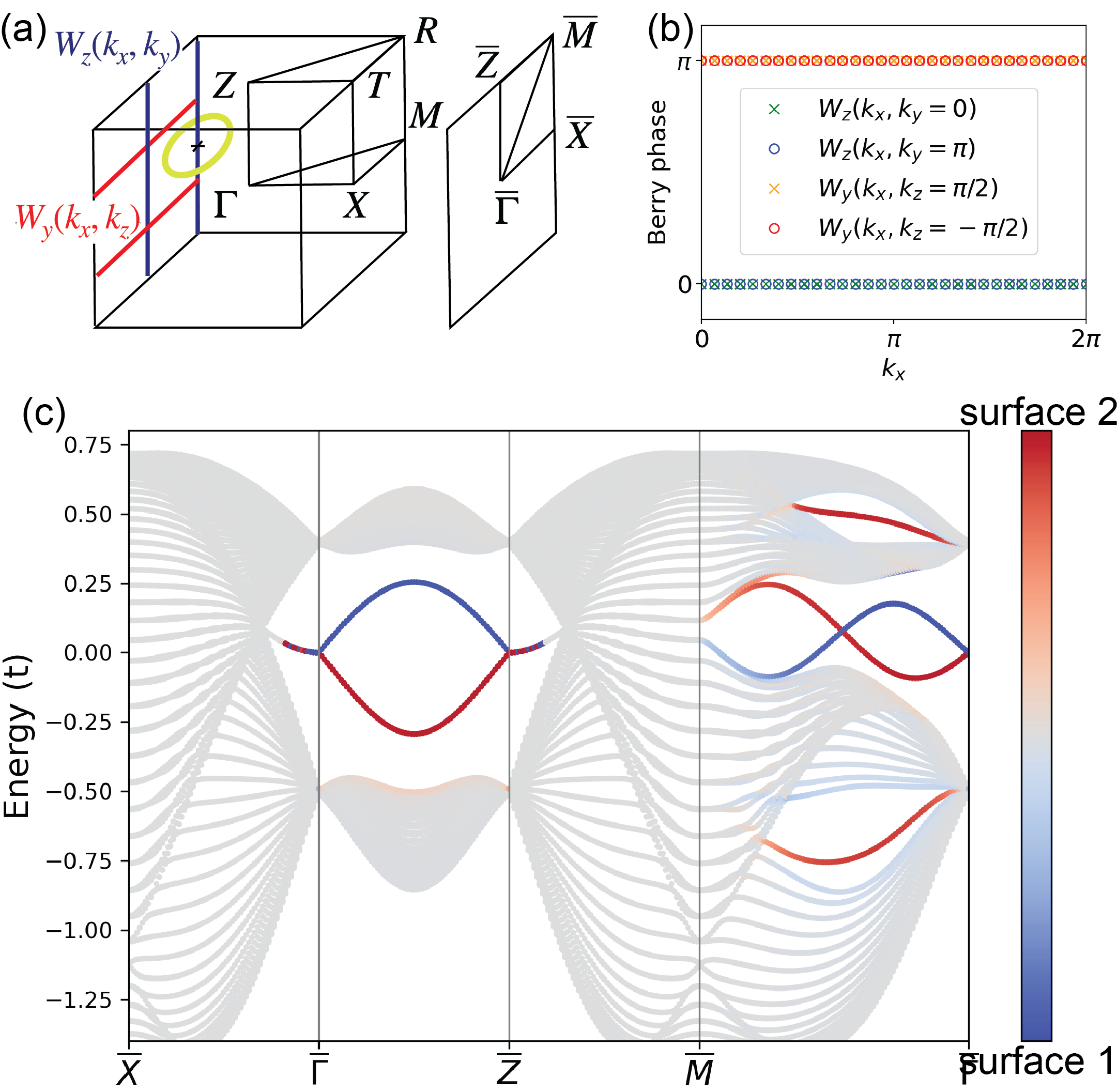}
    \caption{(a) Bulk and surface Brillouin zones. The directions of Wilson loops $W_z(k_x,k_y)$ and $W_y(k_x,k_z)$ are depicted. (b) Wilson loops $W_z(k_x,k_y)$ at $k_y=0,\pi$ and $W_y(k_x, k_z)$ at $k_z=\pm\pi/2$. The nonzero Berry phase $W_y(k_x, k_z)=\pi$ indicates polarization $p_x=p_y=1/2$ at the $k_z$ planes with $k_z\neq 0,\pi$. (c) Surface states with a slab geometry. The colored states are localized at the two boundaries. The presence of the drumhead surface states along $\overline{\Gamma}\overline{Z}$ are results of the quantized polarizations.}
    \label{fig:surface_states}
\end{figure}

The lower two bands below the Fermi energy are gapped except at the HWNLs. This allows us to calculate the Wilson loop of these two bands. We compute the Wilson loop along $y$ directions for $k_x\in[0,2\pi]$, $k_z=\pm\pi/2$, and the Wilson loop along $z$ directions for $k_x\in[0,2\pi]$, $k_y=0,\pi$ as shown in Fig.~\ref{fig:surface_states}(a).
The results, shown in Fig.~\ref{fig:surface_states}(b), reveals a $\pi$ Berry phase for the two bands. 
This Berry phase indicates the quantized polarization $p_x=p_y=1/2 \mod 1$ in unit of $|e|a$ where $e$ is the electron charge and $a$ is the lattice constant.
As a consequence of the Berry phase, there are mid gap surface states at the Fermi energy that are localized on the boundaries of finite systems. These states are shown in Fig.~\ref{fig:surface_states}(c) with red and blue colors representing the two boundaries.

\begin{figure}[ht]
    \centering
    \includegraphics[width=0.6\linewidth]{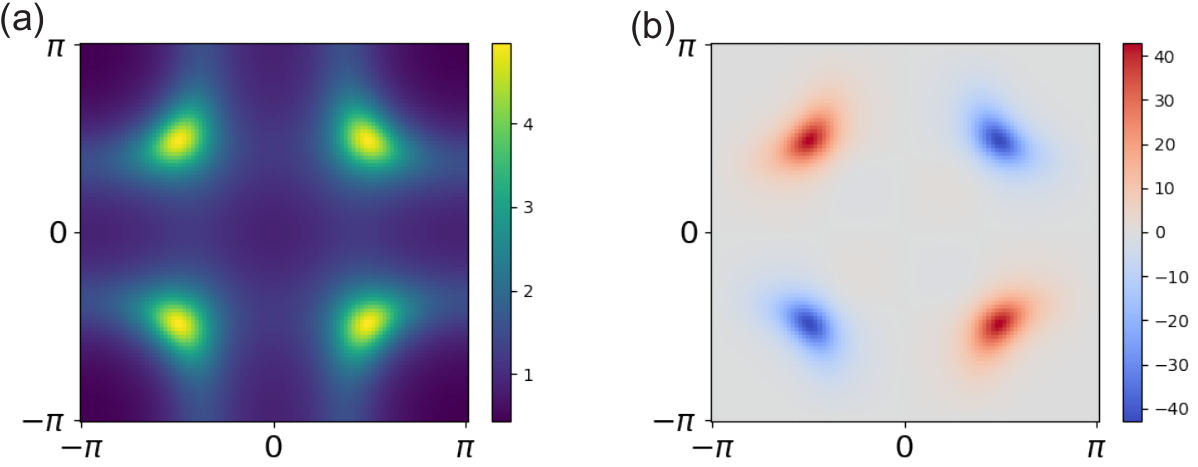}
    \caption{(a) HWNLs and (b) Berry curvature distribution at $k_z=\pi/2$.}
    \label{fig:FS_BC_2}
\end{figure}

The absence of HWNL in the $k_z=\pi/2$ plane is shown in Fig.~\ref{fig:FS_BC_2}(a).
The anti-crossings at $k_z\neq 0,\pi$ where the gap is small are the hot spots of Berry curvature. The Berry curvature distribution at the $k_z=\pi/2$ plane is shown in Fig.~\ref{fig:FS_BC_2}(b).

The degeneracies at high symmetry points/lines and the momentum dependent SOCs are consequences of the magnetic orderings. When the magnetic ordering vanishes, the symmetry group is MSG no.~129.412 (the grey group, or the SG no.~129 with time-reversal symmetry). Each band is two-fold degenerate due to $\cal PT$ symmetry. There are symmetry enforced Dirac points at $X$, $Y$ and $M$ points, protected by glide symmetries and time-reversal symmetry. Typical band structure is shown in Fig.~\ref{fig:129.412} in the next section.
When magnetic ordering turns on, there are momentum dependent, inter-sublattice SOC terms that break the symmetry to MSG no.~129.415, as shown in Eq.~(\ref{eqn:129.415}) in the maintext.

\begin{figure}[ht]
    \centering
    \includegraphics[width=0.4\linewidth]{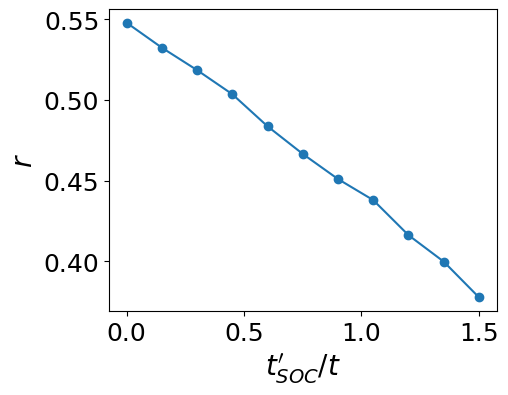}
    \caption{$r$ dependence on the SOC strength.}
    \label{fig:b_vs_M}
\end{figure}

The magnetic ordering in the TM atoms are the sources of the SOC terms $t_{SOC}'$ and $t_{SOC}''$. Here we study the dependence of $r=\langle b\rangle$ with this SOC strength scaled by $t_{SOC}'$.
In this calculation, we choose $t_{SOC}''=2t_{SOC}'$, and fix the energy of the itinerant electron band bottom.
We show that $r$ decreases with the increase of $t_{SOC}'$ in Fig.~\ref{fig:b_vs_M}.
Notice, the magnitude of $t_{SOC}'$ does represent the magnetic moment. In realistic systems, the effective value of $t_{SOC}'$ must be within a range $[0, t_{cutoff}]$. The values depend on the microscopic model of the full system.

\section{Model for $\rm USbTe$ and $\rm CeCoPO$ in MSG no.~129.417}
$\rm USbTe$ and $\rm CeCoPO$ have FM order with MSG no.~129.417 ($P4/nm'm'$). In this section, we construct a toy model to capture the physics of these two materials.

This MSG has generators: four-fold rotation symmetry $C_4$ about the $(\frac14,\frac14,z)$ axis, inversion symmetry $\cal P$ centered at the origin, and two-fold screw symmetry product time-reversal symmetry $\{C_{2x}|\frac12 0 0\}\cal T$. In the basis of $s$-orbitals with spin-orbit coupling at $2a$ Wyckoff position, they are represented by
\begin{align}
    {\cal P} &= \sigma_0\tau_x \\
    {C_{4}} &= \frac{\sigma_0 - \sigma_z}{\sqrt{2}}\tau_0 \\
    {C_{2x}}{\cal T} &= \sigma_z \tau_x {\hat{\cal K}}
\end{align}
where ${\hat{\cal K}}$ is the complex conjugation operator, $\sigma$ and $\tau$ are Pauli matrices that describe spin degrees and sublattice degrees, respectively.

What we consider is an eight-band model constructed by stacking two copies of a four-band model for $s$-orbital on $2a$ Wyckoff position. The Hamiltonian of this eight-band model is denoted as $H_c^{(2)}({\mathbf k})$ and the Hamiltonian of this four-band model is denoted as $H_c^{(1)}({\mathbf k})$. 
The explicit Hamiltonian is following
\begin{align}
    H_c^{(2)}({\mathbf k}) = \begin{pmatrix}
        H_c^{(1)} & m {\mathbf 1}_4 \\
        m {\mathbf 1}_4 & H_c^{(1)}
    \end{pmatrix}
    \label{eqn:g129.417}
\end{align}
where $m$ is the mass term that hybridizes the two four-band models. Here $H_c^{(1)}$ is the Hamiltonian of the four-band model 
\begin{align}
    H_c^{(1)}({\mathbf k}) &= \sum_k \Psi_k^\dagger \left( h_0({\mathbf k}) + h_1({\mathbf k}) - \mu \right) \Psi_k\\
    h_0({\mathbf k}) =& t \cos(\frac{k_x}{2})\cos(\frac{k_y}{2}) \sigma_0\tau_x 
    +t_{SOC} \big( \sin k_x\sigma_y+\sin k_y\sigma_x \big) \tau_z  \\
    h_1({\mathbf k}) &= t_{SOC}'' \left( \cos(\frac{k_y}{2})\sin(\frac{k_x}{2}) \sigma_x - \sin(\frac{k_y}{2})\cos(\frac{k_x}{2}) \sigma_y \right) \tau_y \nonumber \\
     & + t_{SOC}''' \big( \cos k_x+\cos k_y \big) \sigma_z\tau_0  \nonumber \\
     & + t_{SOC}^z \sin(k_z) \left( \cos(\frac{k_y}{2})\sin(\frac{k_x}{2}) \sigma_x - \sin(\frac{k_y}{2})\cos(\frac{k_x}{2}) \sigma_y  \right) \tau_0
\end{align}
where $h_0({\mathbf k})$ preserves $P4/nmm1'$ and $h_1({\mathbf k})$ contains the spin-orbit coupling terms that break $P4/nmm1'$ into MSG no.~129.417 $P4/nm'm'$.
Then we add $f$ electrons using Eq.~(\ref{eqn:model}) and the self-consistent saddle point equations are the same as Eq.~(\ref{eqn:self1}) and Eq.~(\ref{eqn:self2}).

\begin{figure}[t!]
    \centering
    \includegraphics[width=0.8\linewidth]{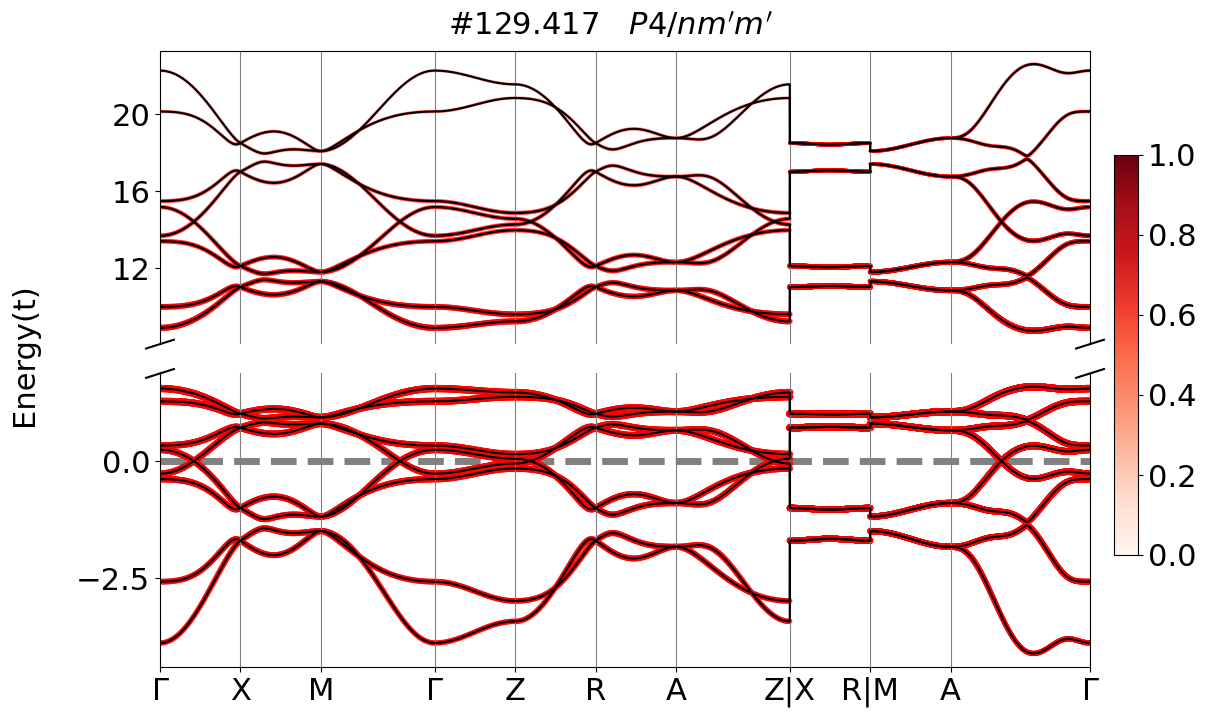}
    \caption{Energy dispersion of the Kondo-driven states in the model (Eq.~(\ref{eqn:g129.417})) along high symmetry lines in the three-dimensional Brillouin zone.}
    \label{fig:g129.417}
\end{figure}
The parameters we choose for this model are: $t=1$, $t_{SOC}=0.5$, $t_{SOC}''=0.2$, $t_{SOC}'''=0$, $t_{SOC}^z=0.5$, $m=4.2$, $V=10$, $\epsilon_d=-4$.
The saddle point solution of this model is shown in Fig.~\ref{fig:g129.417}. The energy scale is scaled with the nearest hopping amplitude $t$.
The Fermi energy crosses Weyl nodal lines as a consequence of the filling constraint of $f$ electrons with strong interactions.
The self-consistent variables are $\lambda=3.666$, $r=0.627$, $\mu=-10.747$.

\begin{figure}[ht]
    \centering
    \includegraphics[width=0.7\linewidth]{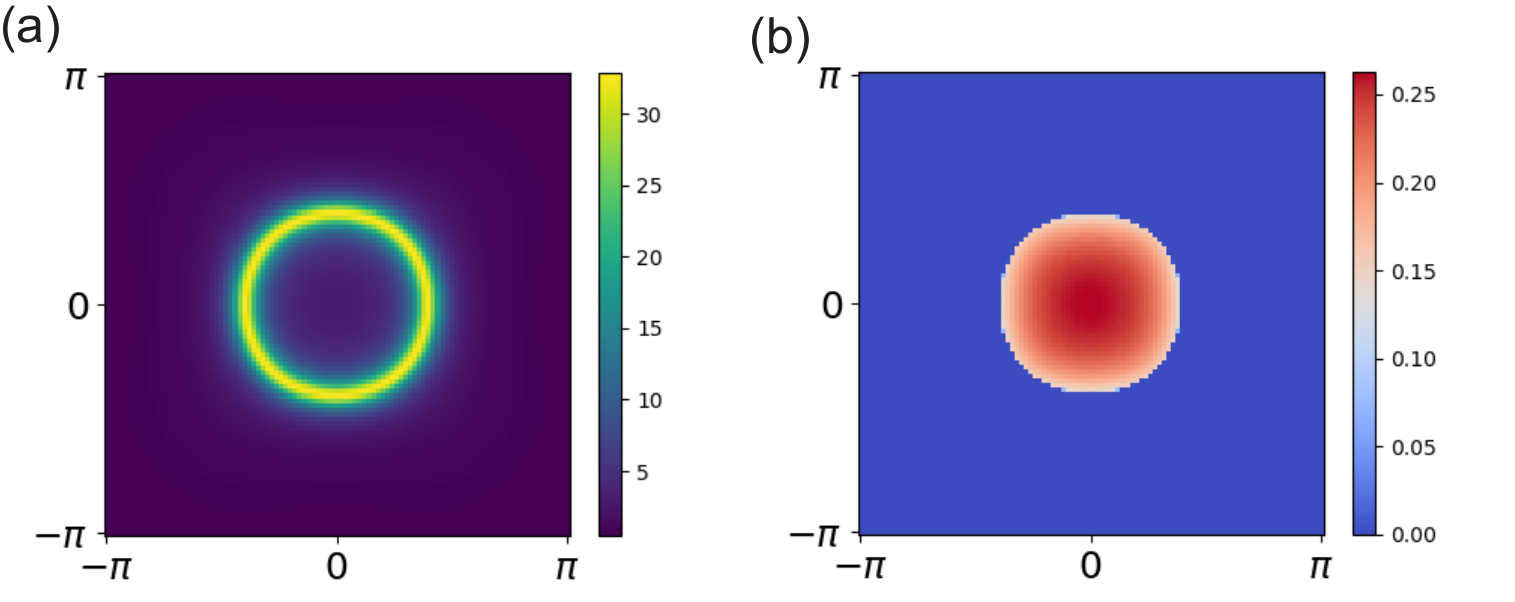}
    \caption{(a) The Weyl nodal lines at Fermi energy for the model Eq.~(\ref{eqn:g129.417}). (b) Berry curvature of the lowest four bands. }
    \label{fig:gBC_WNDL}
\end{figure}
The Weyl nodal lines and Berry curvature of the lowest four bands are shown in Fig.~\ref{fig:gBC_WNDL}. The Berry curvature shows there is first order anomalous Hall effect since the Fermi surface encloses the pocket in the Brillouin zone center where the Berry curvature is large.

\begin{figure}[ht]
    \centering
    \includegraphics[width=0.75\linewidth]{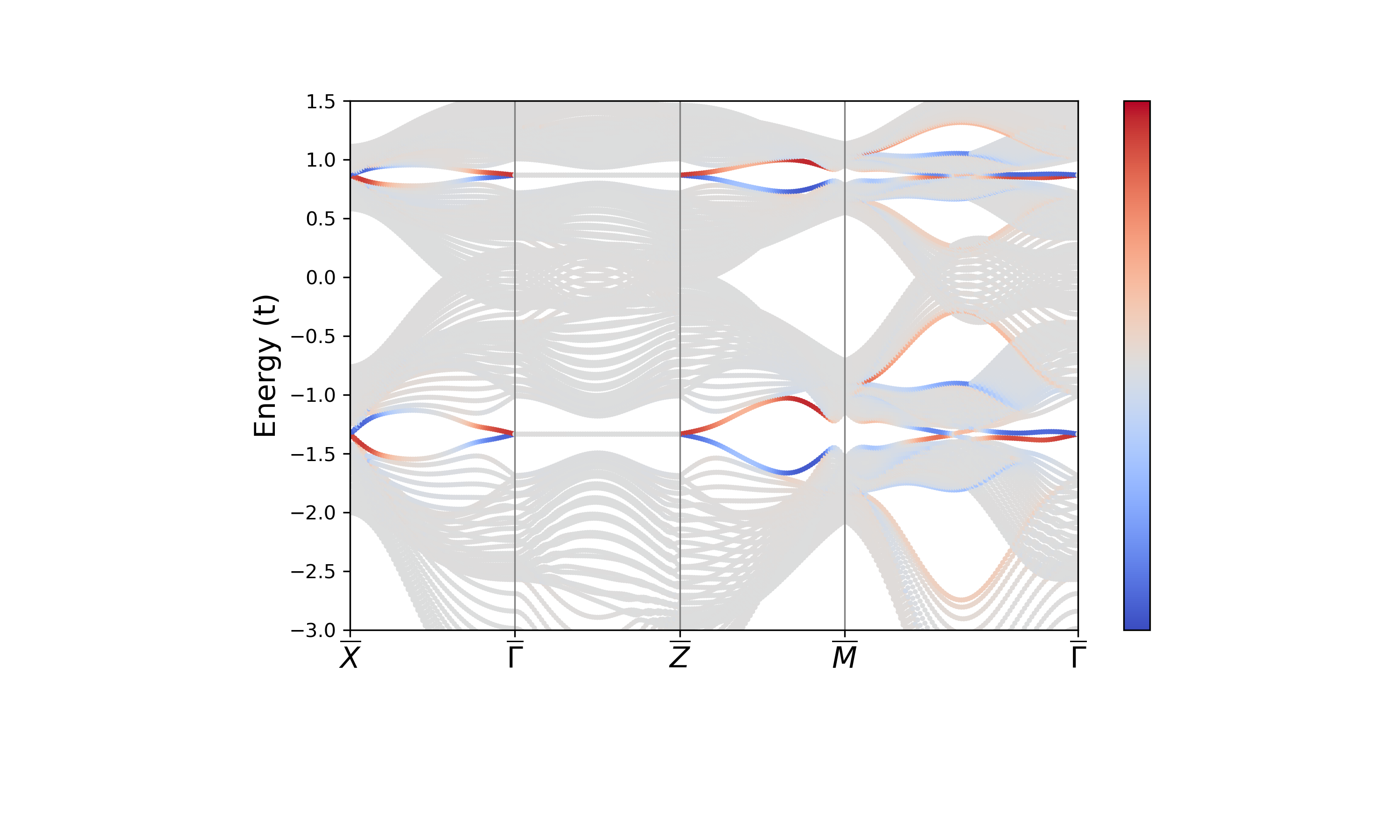}
    \caption{The surface states of the $f$ bands around Fermi energy for the model Eq.~(\ref{eqn:g129.417}).}
    \label{fig:gsurface}
\end{figure}
We also show the surface states of the $f$ bands near Fermi energy in Fig.~\ref{fig:gsurface}. 
The surface states are calculated on a slab geometry with termination along $x$ direction.
There are topological surface states in the two gaps around $1t$ and $-1.5t$. However, the surface states of the Weyl nodal lines are obscured by the projections of bulk states.

\section{DFT calculation for $\rm CeCoPO$}
\begin{figure}
    \centering
    \includegraphics[width=0.6\linewidth]{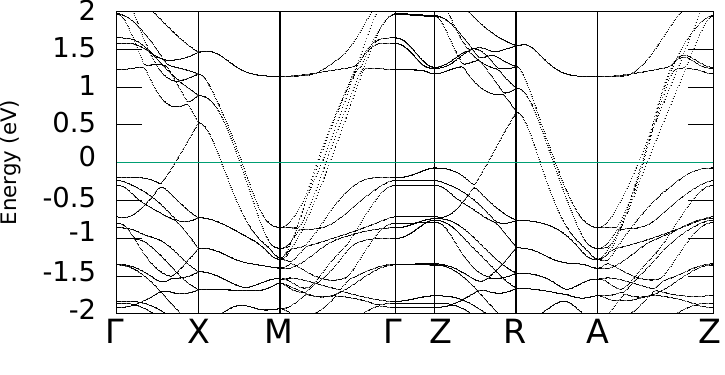}
    \caption{DFT band structure of $\rm CeCoPO$.}
    \label{fig:DFT}
\end{figure}
\begin{figure}[ht]
    \centering
    \includegraphics[width=0.9\linewidth]{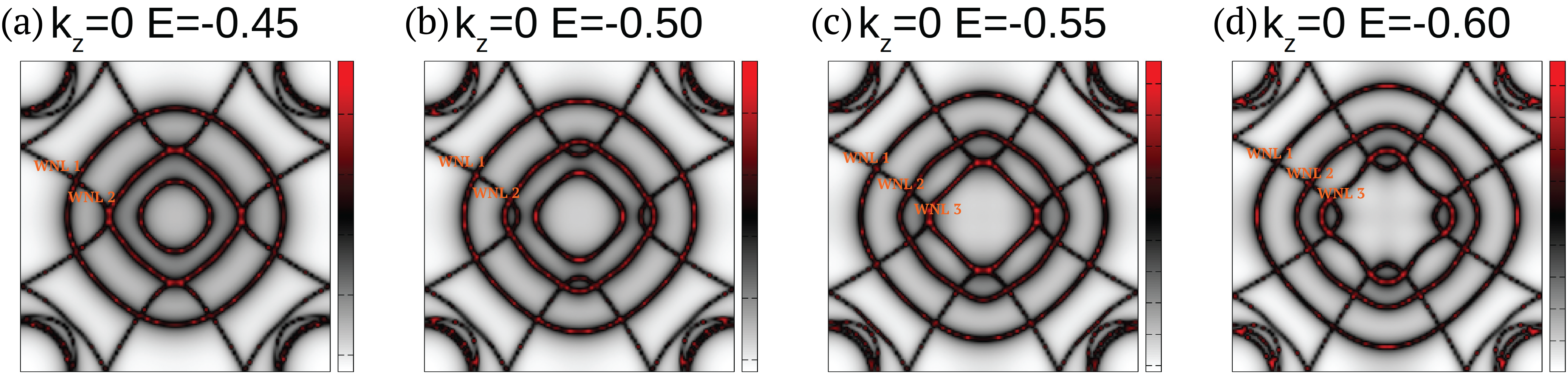}
    \caption{WNLs in the $k_z=0$ plane at different energy are marked with orange texts. The crossing between two band pockets moves as energy changes, showing it is a loop in the $k_x-k_y-E$ space. }
    \label{fig:WNL_DFT}
\end{figure}

We performed density functional theory (DFT) calculations to study the electronic structure of $\rm CeCoPO$ as implemented in the full-potential local-orbital (FPLO) code \cite{FPLO1,FPLO2}. We treat the $4f$ electrons of Ce as core electrons and we remove them from the local orbital basis to solve the Schr\"odinger equation. We use the meta-Becke-Johnson exchange functional (MBJ) \cite{MBJ} within the meta-generalized gradient approximation (meta-GGA). We used a Monkhorst-Pack grid of $7\times7\times5$ k-points for reciprocal space integration. Within this approach, we obtain a magnetization of 0.66$\mu_{B}$ per Co atom. To analyze the Fermi surface cuts we construct maximally localized Wannier functions for the valence orbitals as implemented in FPLO, choosing as basis $\{6s, 5d, 6d, 6p, 5f\}$ orbitals for Ce, $\{4s, 3d, 4d, 4p\}$ orbitals for Co, $\{2p, 3p\}$ orbitals for O and $\{3s, 3p, 4p, 3d\}$ orbitals for P.


The DFT bands show degeneracies along $XM$ and $RA$, which are consistent with the symmetry analsis.
There are also Weyl nodal lines in the $k_z=0,\pi$ plane. Those are symmetry protected nodal lines. 
Fig.~\ref{fig:WNL_DFT} shows the band crossings in the $k_z=0$ plane for different energy. The crossings move as energy changes, showing that they are closed loops in the $k_x-k_y-E$ space, i.e. they are WNLs.
For example, the crossing along $\Gamma X$ at $E=-0.45$ is seen in both the DFT band structure Fig.~\ref{fig:DFT} and the constant energy contour Fig.~\ref{fig:WNL_DFT}.

\section{\label{app:examples}Examples in MSG 129 family and MSG 125.367}
\begin{figure}[ht]
    \centering
    \includegraphics[width=0.5\linewidth]{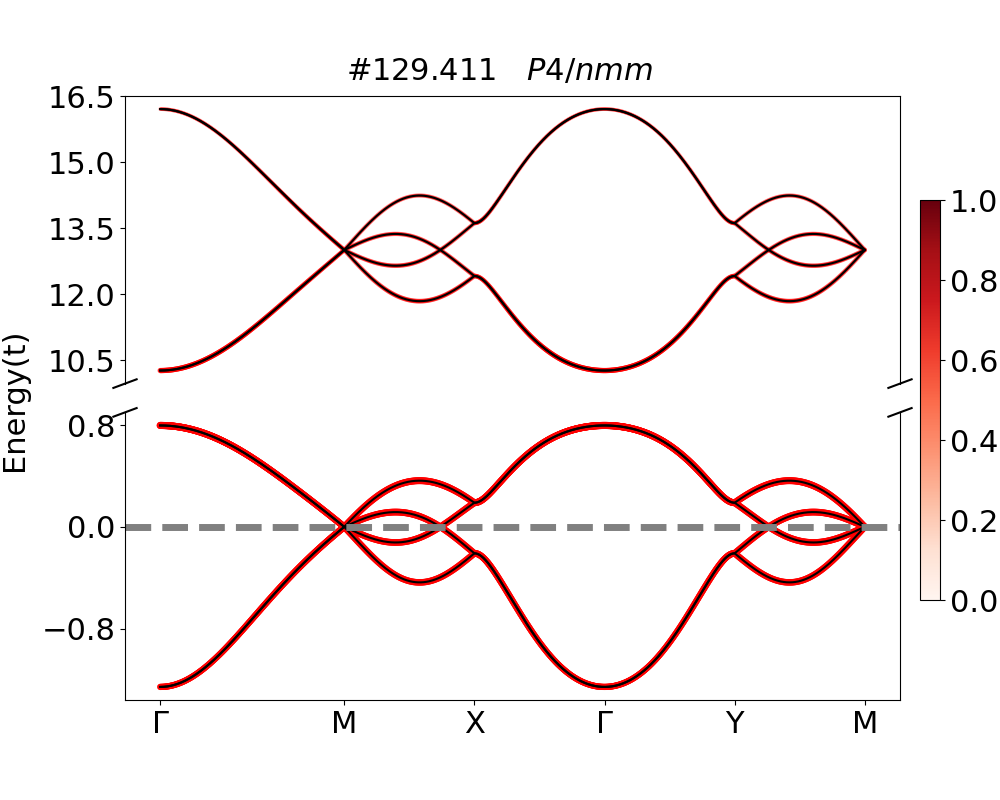}
    \caption{
    Energy dispersion of the model Eq.~(\ref{eqn:129.411}) in MSG no.~129.411 ($P4/nmm$). Red color indicates the ratio of $f$ electrons in the states.}
    \label{fig:129.411}
\end{figure}

\begin{figure}[ht]
    \centering
    \includegraphics[width=0.5\linewidth]{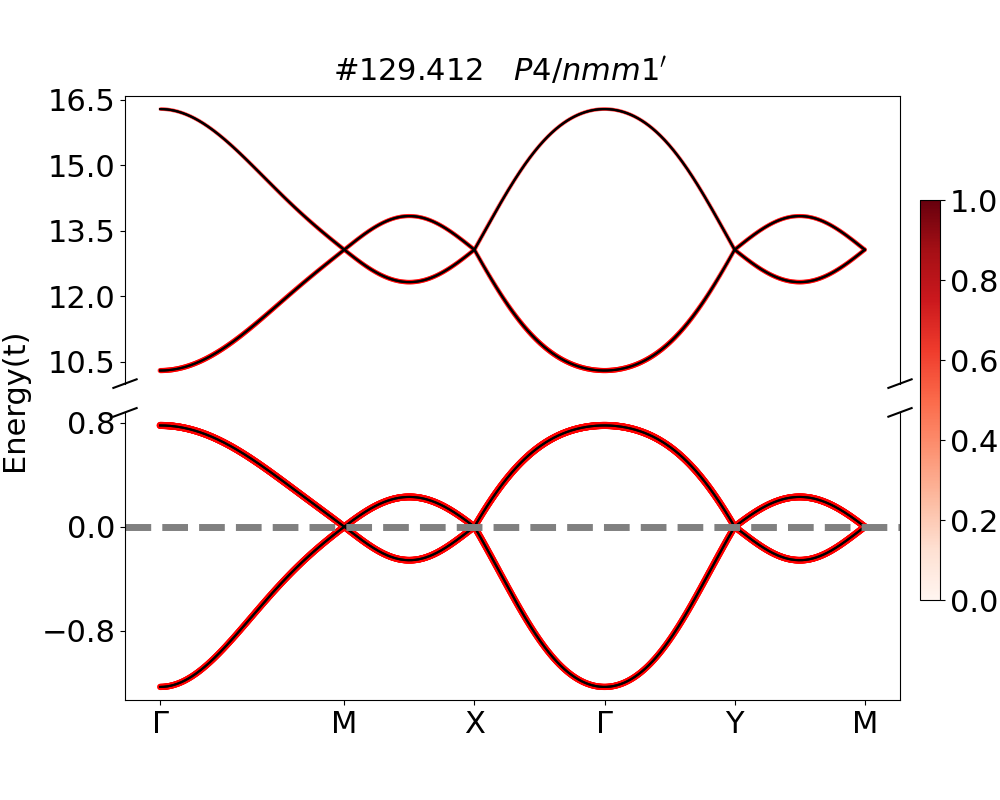}
    \caption{
    Energy dispersion of the model Eq.~(\ref{eqn:129.412}) in MSG no.~129.412 ($P4/nmm1'$). }
    \label{fig:129.412}
\end{figure}

\begin{figure}[ht]
    \centering
    \includegraphics[width=0.5\linewidth]{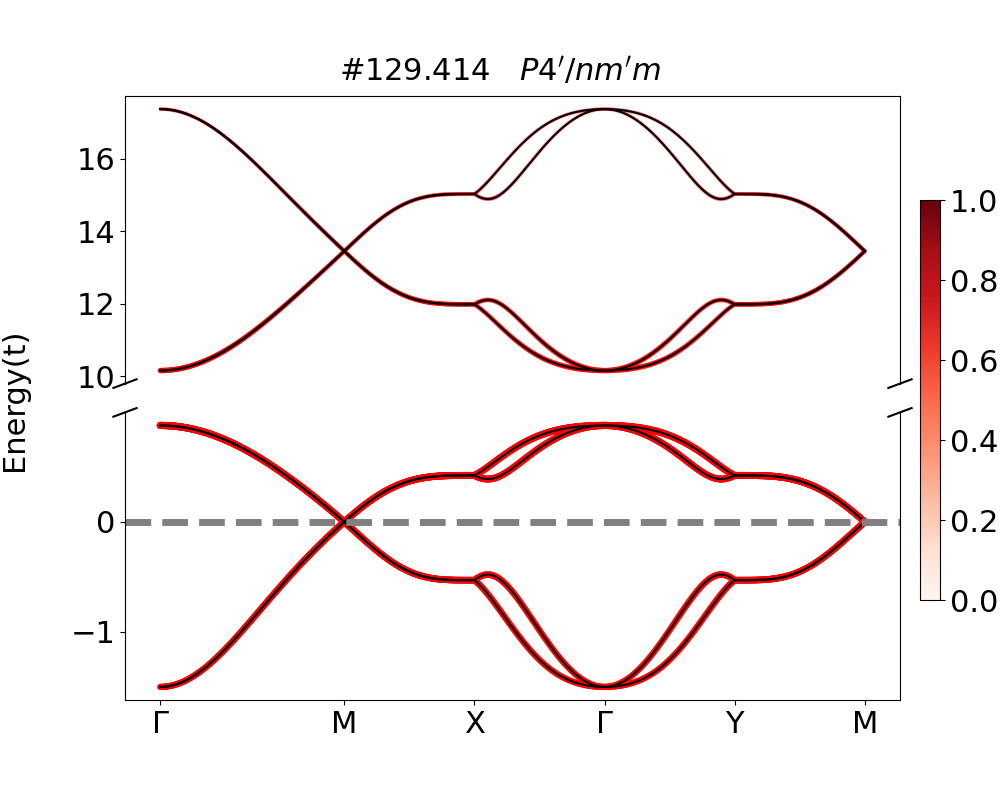}
    \caption{Band structure of the model Eq.~(\ref{eqn:129.414}) in MSG no.~129.411 ($P4'/nm'm$).}
    \label{fig:129.414}
\end{figure}

\begin{figure}[ht]
    \centering
    \includegraphics[width=0.5\linewidth]{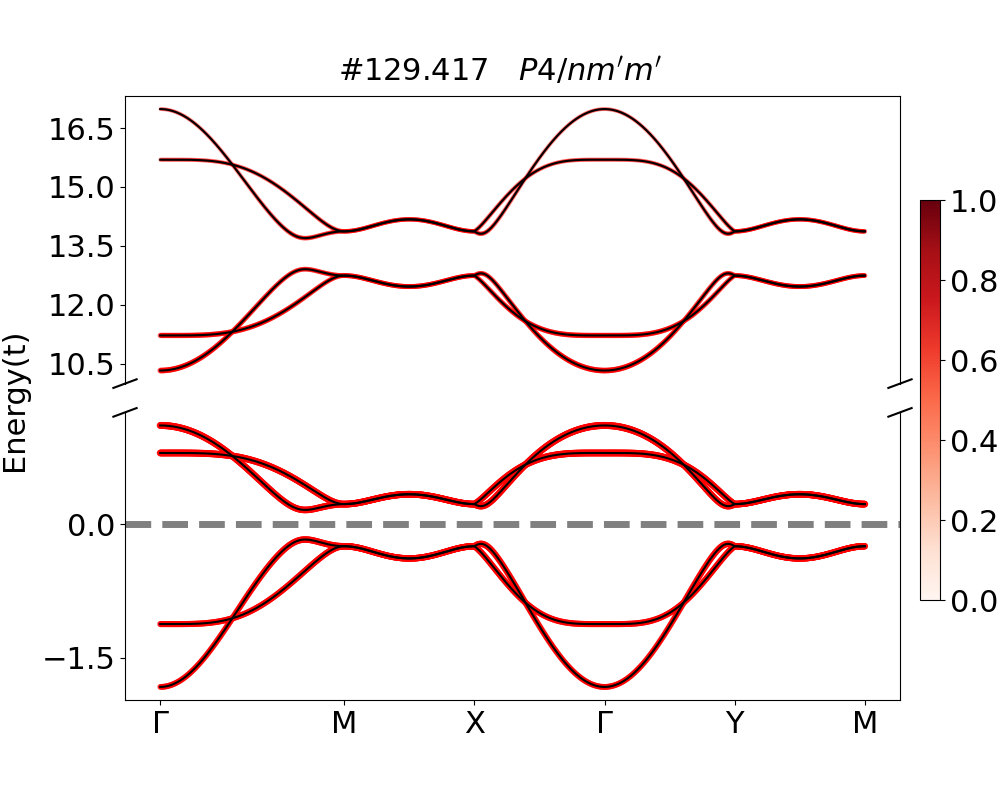}
    \caption{Band structure of the model Eq.~(\ref{eqn:129.417}) in MSG no.~129.417 ($P4/nm'm'$).}
    \label{fig:129.417}
\end{figure}

In this section we first present additional examples in the MSG 129 family.
The symmetries of these groups are a subset of the grey group $P4/nmm1'$. For this grey group the generators are time-reversal symmetry $\cal T$, four-fold rotation symmetry $C_4$ about the $(\frac14,\frac14,z)$ axis, inversion symmetry $\cal P$ centered at the origin, and two-fold screw symmetry $\{C_{2x}|\frac12 0 0\}$. In the basis of $s$-orbitals with spin-orbit coupling at $2a$ Wyckoff position, they are represented by
\begin{align}
    {\cal T} &= -i\sigma_y \tau_0 {\hat{\cal K}} \\
    {\cal P} &= \sigma_0\tau_x \\
    {C_{4}} &= \frac{\sigma_0 - \sigma_z}{\sqrt{2}}\tau_0 \\
    {C_{2x}} &= -i\sigma_x \tau_x
\end{align}
where ${\hat{\cal K}}$ is the complex conjugation operator, $\sigma$ and $\tau$ are Pauli matrices that describe spin degrees and sublattice degrees, respectively.
The MSGs we study in this section include no.~129.411 ($P4/nmm$), no.~129.412 ($P4/nmm1'$), no.~129.414 ($P4'/nm'm$) and no.~129.417 ($P4/nm'm'$). The compatible magnetic ordering and nonlinear Hall responses are listed in Table~\ref{tab:tab129} in the main text. 
The itinerant electron Hamiltonian of these groups are of the form
\begin{align}
    \label{eqn:app_Hc}
    H_c({\mathbf k}) &= \sum_k \Psi_k^\dagger \left( h_0({\mathbf k}) + h_1({\mathbf k}) - \mu \right) \Psi_k\\
    h_0({\mathbf k}) =& t \cos(\frac{k_x}{2})\cos(\frac{k_y}{2}) \sigma_0\tau_x 
    +t_{SOC} \big( \sin k_x\sigma_y+\sin k_y\sigma_x \big) \tau_z  
    \label{eqn:app_h0}
\end{align}
and $h_1({\mathbf k})$ contains the spin-orbit coupling terms that break $P4/nmm1'$ into the desired MSGs. 
The band structures of these models are shown in Figs.~\ref{fig:129.411},~\ref{fig:129.412},~\ref{fig:129.414} and \ref{fig:129.417}. The different types of spin splittings can be seen from these plots. The spin splittings can be checked by studying the symmetries at those momenta.
The types of symmetry enforced crossings agree with Table~\ref{tab:tab129} in the main text. 


The corresponding spin-orbit coupling terms $h_1({\mathbf k})$, with the aforementioned groups labeled by the MSGs in order,  are as follows
\begin{align}
    h_1^{(129.411)}({\mathbf k}) & = t_{SOC}'' \left( \cos(\frac{k_x}{2})\sin(\frac{k_y}{2}) \sigma_x + \sin(\frac{k_x}{2})\cos(\frac{k_y}{2}) \sigma_y \right) \tau_y \label{eqn:129.411}\\
    h_1^{(129.412)}({\mathbf k}) & =0 \label{eqn:129.412}\\
    h_1^{(129.414)}({\mathbf k}) & = +t_{SOC}' \sin(\frac{k_x}{2})\sin(\frac{k_y}{2}) \sigma_z\tau_x  \nonumber\\
    &+t_{SOC}'' \left( \cos(\frac{k_x}{2})\sin(\frac{k_y}{2}) \sigma_x - \sin(\frac{k_x}{2})\cos(\frac{k_y}{2}) \sigma_y \right) \tau_y \label{eqn:129.414}\\
    h_1^{(129.417)}({\mathbf k})
    &= t_{SOC}'' \left( \cos(\frac{k_y}{2})\sin(\frac{k_x}{2}) \sigma_x - \sin(\frac{k_y}{2})\cos(\frac{k_x}{2}) \sigma_y \right) \tau_y \nonumber \\
     & + t_{SOC}''' \big( \cos k_x+\cos k_y \big) \sigma_z\tau_0  \nonumber \\
     & + t_{SOC}^z \sin(k_z) \left( \cos(\frac{k_y}{2})\sin(\frac{k_x}{2}) \sigma_x - \sin(\frac{k_y}{2})\cos(\frac{k_x}{2}) \sigma_y  \right) \tau_0
     \label{eqn:129.417}
\end{align}

\begin{figure}[ht]
    \centering
    \includegraphics[width=0.5\linewidth]{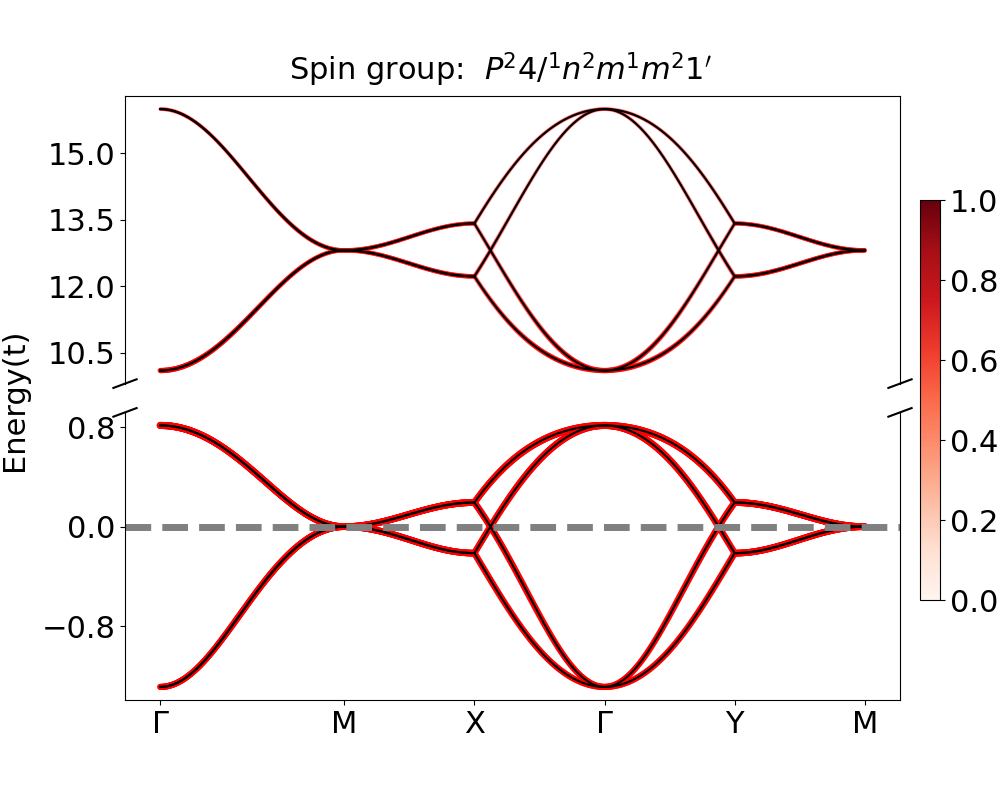}
    \caption{
    Energy dispersion of the model Eq.~(\ref{eqn:spin_group}) in spin group $P{}^24/{}^1n{}^1m{}^2m{}^21'$.}
    \label{fig:spin_group}
\end{figure}

MSG no.~129.415 ($P4'/nmm'$) and MSG no.~125.367 ($P4'/nbm'$) have many symmetries in common. They both have $C_4{\cal T}$ rotation about $(\frac14,\frac14,z)$ axis and inversion symmetry $\cal P$ centered at the origin. The difference is that $P4'/nmm'$ has screw symmetry $\{C_{2x}|\frac12 00\}$ while $P4'/nbm'$ has the glide symmetry $\{C_{2x}|0\frac12 0\} {\cal P}$.
In the basis of $s$-orbitals in $2a$ Wyckoff position, the matrix representation is $\{C_{2x}|0\frac12 0\}= -i\sigma_x\tau_0$. 
The Hamiltonian of MSG no.~125.367 ($P4'/nbm'$) is
\begin{align}
    h^{(125.367)}({\mathbf k}) =& t \cos(\frac{k_x}{2})\cos(\frac{k_y}{2}) \sigma_0\tau_x + t_z \sin k_z\cos(\frac{k_x}{2})\cos(\frac{k_y}{2}) \sigma_0\tau_y \nonumber\\
    &+t_{SOC} \big( \sin k_x\sigma_x-\sin k_y\sigma_y \big) \tau_z   \nonumber\\
    &+t_z' \sin k_z \left( \cos k_x - \cos k_y \right) \sigma_z \tau_z
    +t_{SOC}' \sin(\frac{k_x}{2})\sin(\frac{k_y}{2}) \sigma_z\tau_x 
    \nonumber\\
    &+t_{SOC}' \sin(\frac{k_x}{2})\sin(\frac{k_y}{2}) \sigma_z\tau_x  \nonumber\\
    &+t_{SOC}'' \left( \cos(\frac{k_x}{2})\sin(\frac{k_y}{2}) \sigma_y + \sin(\frac{k_x}{2})\cos(\frac{k_y}{2}) \sigma_x \right) \tau_y
\end{align}
For comparison, we also list the Hamiltonian of MSG no.~129.415 ($P4'/nmm'$) here (Eq.~(\ref{eqn:Hc}) in the main text)
\begin{align}
    h^{(129.415)}({\mathbf k}) &= t \cos(\frac{k_x}{2})\cos(\frac{k_y}{2}) \sigma_0\tau_x + t_z \sin k_z\cos(\frac{k_x}{2})\cos(\frac{k_y}{2}) \sigma_0\tau_y \nonumber\\
    &+t_{SOC} \big( \sin k_x\sigma_y+\sin k_y\sigma_x \big) \tau_z  \nonumber\\
    &+t_z' \sin k_z \left( \cos k_x - \cos k_y \right) \sigma_0 \tau_z
    +t_{SOC}' \sin(\frac{k_x}{2})\sin(\frac{k_y}{2}) \sigma_z\tau_x  \nonumber\\
    &+t_{SOC}'' \left( \cos(\frac{k_x}{2})\sin(\frac{k_y}{2}) \sigma_x - \sin(\frac{k_x}{2})\cos(\frac{k_y}{2}) \sigma_y \right) \tau_y 
\end{align}
Therefore, when the parameters are chosen equally, the band structures (Fig.~\ref{fig:129.415}) and the Berry curvature distributions (Fig.~\ref{fig:FS_BC}) of the two models are identical.

We also present one example of the spin space group (for an introduction of spin space group, see Ref.~\cite{litvin1974spin}) $P{}^24/{}^1n{}^1m{}^2m{}^21'$ in Fig.~\ref{fig:spin_group}. It is generated by spin group symmetries $[C_2||C_4]$, $[E||{\cal P}]$, $[C_2||\{C_{2x}|\frac12 00\}]$, and $[C_2||E]{\cal T}$, where $[A||B]$ indicates the symmetry is composed of a symmetry $A$ acting in the spin space and a symmetry $B$ acting in the real space. The Hamiltonian of this group is
\begin{align}
    h^{\text{spin group}}({\mathbf k}) =& t \cos(\frac{k_x}{2})\cos(\frac{k_y}{2}) \sigma_0\tau_x + t_z \sin k_z\cos(\frac{k_x}{2})\cos(\frac{k_y}{2}) \sigma_0\tau_y \nonumber\\
    &+ \big( \cos k_x - \cos k_y\big) \sigma_y \big(t_{SOC1}\tau_z + t_{SOC2} \tau_x\big) \label{eqn:spin_group}
\end{align}
The band structure of this model is shown in Fig.~\ref{fig:spin_group}, where the symmetry enforced Dirac point and HWNLs coexist.

\section{\label{app:tables} MSGs in SG no.~125 and SG no.~31 families}
\begin{table}[ht]
    \centering
    \begin{tabular}{cc|c|c|c |c}
         MSG \#&BNS  & FM & AFM & Hall response &Crossing type\\
         \hline 
         125.363 &$P4/nbm$ & &\checkmark & 3rd- order & Dirac\\
         125.364 &$P4/nbm1'$ & & &  & Dirac\\
         125.365 &$P4/n'bm$ & &\checkmark &  & \\
         125.366 &$P4'/nb'm$ & &\checkmark & 3rd- order & Dirac\\
         125.367 &$P4'/nbm'$ & &\checkmark & 3rd- order & HWNL\\
         125.368 &$P4'/n'b'm$ & &\checkmark & &Dirac  \\
         125.369 &$P4/nb'm'$ & \checkmark &\checkmark & 1st-order  &  \\
         125.370 &$P4'/n'bm'$ & &\checkmark & & HWNL \\
         125.371 &$P4/n'b'm'$ & &\checkmark & & HWNL  \\
         \hline
         125.372 &$P_{c}4/nbm$ & &\checkmark & & Dirac\\
         125.373 &$P_{C}4/nbm$ & &\checkmark &  & Dirac\\
         125.374 &$P_{I}4/nbm$ & &\checkmark &  & double Dirac\\
    \end{tabular}
    \caption{MSGs of the space group no.~125 family. We use BNS setting to denote the MSGs. 
    The magnetic orderings compatible with each MSG are marked with $\checkmark$. Blank means the magnetic ordering is not compatible with the group. The leading order Hall response of each group is listed. Blank means no Hall response of any order is permitted.
    The crossing types for bands from $s$-orbitals at $2a$ Wyckoff positions are listed. Blank means no symmetry-enforced Dirac point, hourglass Weyl nodal line, or double Dirac point exists in the group.
    }
    \label{tab:tab125}
\end{table}
\begin{table}[ht]
     \centering
    \begin{tabular}{cc|c|c|c |c}
         MSG \#&BNS  & FM & AFM & Hall response &Crossing type\\
         \hline 
         31.123 & $Pmn2_1$ & &\checkmark &2nd-order & \\
         31.124 & $Pmn2_11'$ & & &2nd-order& Dirac\\
         31.125 & $Pm'n2_1'$ &\checkmark&\checkmark& 1st-order & \\
         31.126 & $Pmn'2_1'$ &\checkmark&\checkmark& 1st-order & \\
         31.127 & $Pm'n'2_1$ &\checkmark&\checkmark&1st-order  & \\
         \hline
         31.128 & $P_amn2_1$ & &\checkmark &2nd-order & Dirac \\
         31.129 & $P_bmn2_1$ & &\checkmark &2nd-order & Dirac \\
         31.130 & $P_cmn2_1$ & &\checkmark &2nd-order & HWNL\\
         31.131 & $P_Amn2_1$ & &\checkmark &2nd-order & HWNL \\
         31.132 & $P_Bmn2_1$ & &\checkmark &2nd-order & \\
         31.133 & $P_Cmn2_1$ & &\checkmark &2nd-order & Dirac \\
         31.134 & $P_Imn2_1$ & &\checkmark &2nd-order & 
    \end{tabular}
    \caption{MSGs of the space group no.~31 family. The interpretation of this table is the same as for Table~\ref{tab:tab129}.
    }
    \label{tab:tab31}
\end{table}

The SG 125 family is very similar to the SG 129 family. As we have discussed in the previous section, they differ by the two-fold rotation symmetry about $x$ and $y$-axis. Therefor, the compatible magnetic ordering and the corresponding Hall responses are identical in Table~\ref{tab:tab129} and Table~\ref{tab:tab125}. However, the symmetry enforced crossings can be different as we listed in Table~\ref{tab:tab125}. In particular, in MSG no.~125.374 ($P_{I}4/nbm$) there is a double Dirac point at $(\pi,\pi,\pi)$ due to the interplay between anti-unitary symmetries and nonsymmorphic groups.

For the 31 family, MSGs no.~31.125-31.127 are compatible with a FM ordering while all MSGs in this family except no.~31.124 are compatible with certain AFM ordering. MSGs no.~31.123, no.~31.124 and no.31.128-31.134 have 2nd-order responses as the leading order. 

For MSG no.~31.130 and no.~31.131, the compatibility conditions require the following irreps connectivity: $\overline{R}_2\overline{R}_4(2)\rightarrow \overline{N}_3(1)\oplus \overline{N}_4(1)$, $\overline{R}_3\overline{R}_5(2)\rightarrow \overline{N}_3(1)\oplus \overline{N}_4(1)$, and $\overline{S}_2\overline{S}_5(2)\rightarrow \overline{N}_4(1)\oplus \overline{N}_4(1)$, $\overline{S}_3\overline{S}_4(2)\rightarrow \overline{N}_3(1)\oplus \overline{N}_3(1)$, where $N=(u,\pi,w)$, $R=(\pi,\pi,\pi)$, $S=(\pi,\pi,0)$.
Therefore, there are symmetry enforced HWNLs in the $k_y=\pi$ plane. 

\section{Prime material candidates}

\begin{figure}
    \centering
    \includegraphics[width=0.8\linewidth]{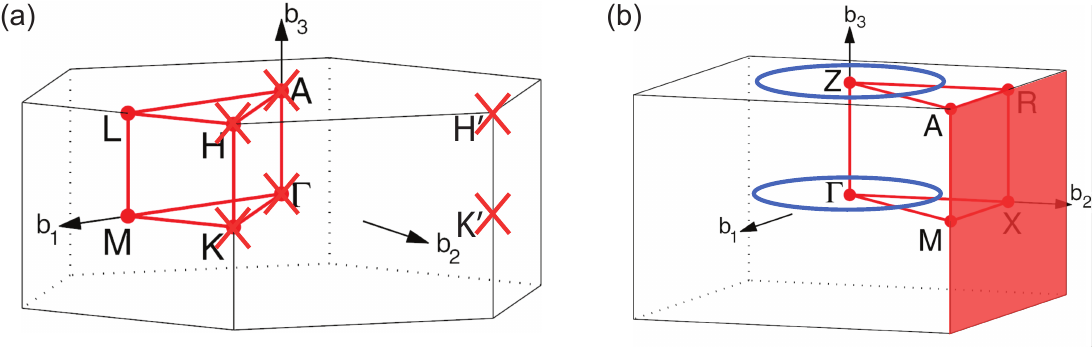}
    \caption{(a) Brillouin zone of MSG no~189.224 ($P\bar 6'2m'$) ($\rm UNiGa$ and $\rm UNiAl$). Red crosses indicate the momenta that may have symmetry enforced Weyl points. (b) Brillouin zone of MSG no.~129.417 ($P4/nm'm'$) ($\rm CeCoPO$ and $\rm USbTe$). The red plane is indicating the states on Brillouin zone boundaries $k_x=\pi$ and $k_y=\pi$ are two-fold degenerate enforced by symmetries. The blue circles indicate the planes $k_z=0$ and $k_z=\pi$ that may have Weyl nodal lines. }
    \label{fig:BZs}
\end{figure}

In this section, we explain further details about the prime candidates that have been discussed in the main texts.

\subsection{$\rm UNiGa$}
We predict $\rm UNiGa$~\cite{UNiGa,UNiGamag} as a MWKSM candidate with a 3rd order Hall response. This material has an AFM order with MSG no~189.224 ($P\bar 6'2m'$). The parent group is SG no~189. The magnetic moment is on the $U$ site with $\mu_{eff}=2.71 \mu_B$ and the ordered moment is $\mu_s=1.30 \mu_B$ per $\rm U$, oriented 
along the z-direction~\cite{UNiGa}. $\rm UNiGa$ has magnetic structure (+ - - + + -). It has $\rm U$ sitting at the $\rm 6i$ Wyckoff position.
The Neel temperatures is 38K. The Sommerfeld coefficients are 59 $\rm mJ/K^2mol$. The resistivity $\rho$ at low temperature is 95 $\rm \mu\Omega cm$ with semimetallic behavior as shown in Fig.~\ref{fig:resistivity} (a). This material has symmetry enforced Weyl points at high symmetry points as Fig.~\ref{fig:BZs} (a) shows.

\subsection{$\rm UNiAl$}
Similar to $\rm UNiGa$, we predict $\rm UNiAl$~\cite{UNiGa,UNiGamag} as a MWKSM candidate with a 3rd order Hall response. This material has an AFM order with MSG no~189.224 ($P\bar 6'2m'$). The parent group is SG no~189. The magnetic moment is on the $U$ site with $\mu_{eff}=1.70 \mu_B$ and is oriented in z-direction~\cite{UNiGa}. $\rm UNiAl$ has magnetic structure (+ - + -). It has $\rm U$ sitting at the $\rm 6i$ Wyckoff position.
The Neel temperatures is 21K. The Sommerfeld coefficients is 160 $\rm mJ/K^2mol$. The resistivity $\rho$ at low temperature is 215 $\rm \mu\Omega cm$ with semimetallic behavior as shown in Fig.~\ref{fig:resistivity} (a). This material has symmetry enforced Weyl points at high symmetry points as Fig.~\ref{fig:BZs} (a) shows.

\subsection{$\rm CeCoPO$}
We predict $\rm CeCoPO$~\cite{CeCoPO} as a candidate of MWKSM with a FM order. 
Clearly it has a 1st order anomalous Hall effect due to the FM order. This material mainly has magnetic moment on the $\rm Co$ sites.
The ordered moment is $\mu_s=0.4 \mu_B$ per $\rm U$ and the effective moment is $\mu_{eff}=2.9 \mu_B$~\cite{CeCoPO}. 
The Curie temperature is 75K.
The Sommerfeld coefficient is $\gamma=200 {\rm mJ/K^2mol}$ (see Fig.~\ref{fig:specific_heat} (b)).
The resistivity is shown in Fig.~\ref{fig:resistivity} (b), where the resisdual resistivity ratio (RRR) is about $10$, but $\rho_{{\rm low}-T} \approx 750 {\rm \mu\Omega cm}$ (and $\rho_{T=300K} \approx 5 {\rm m\Omega cm}$)
The carrier concentration is 
expected to be around $1\times10^{21} {\rm cm}^{-3}$, as estimated from Hall investigations of a comparable system -- the doped $\rm CeFeAsO$~\cite{Yuan_2011,MagneticStudy_CeFeAsOF,Jaroszynski2008}.
Therefore, we predict it as a MWKSM. 
This material has symmetry enforced degeneracy at the Brillouin zone boundary and symmetry protected Weyl nodal lines as 
Fig.~\ref{fig:BZs} (b) shows.

\subsection{$\rm USbTe$}
\begin{figure}
    \centering
    \includegraphics[width=1.0\linewidth]{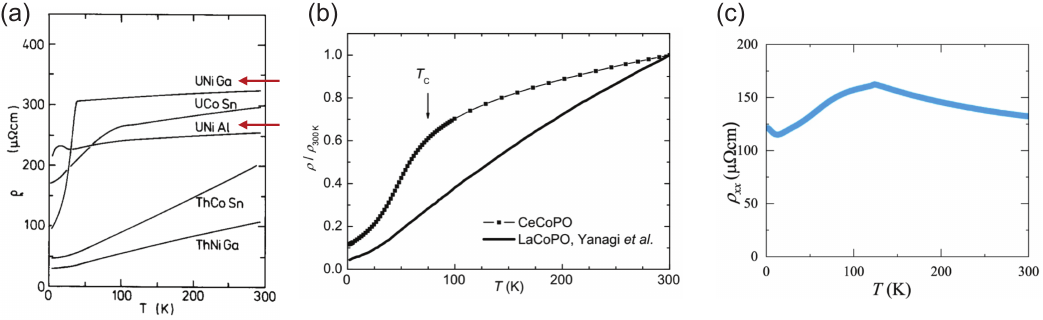}
    \caption{The temperature dependence of resistivity for (a) $\rm UNiGa$ and $\rm UNiAl$,
    from Ref.~\cite{UNiGa}; (b) $\rm CeCoPO$,
    from Ref.~\cite{CeCoPO} and (c) $\rm USbTe$, 
    from Ref.~\cite{USbTe}.
   For CeCoPO, the $\rho_{{\rm low}-T}
   \approx 750 {\rm \mu\Omega cm}$, and $\rho_{T=300K} \approx 5 {\rm m\Omega cm}$.}
    \label{fig:resistivity}
\end{figure}

\begin{figure}
    \centering
    \includegraphics[width=0.7\linewidth]{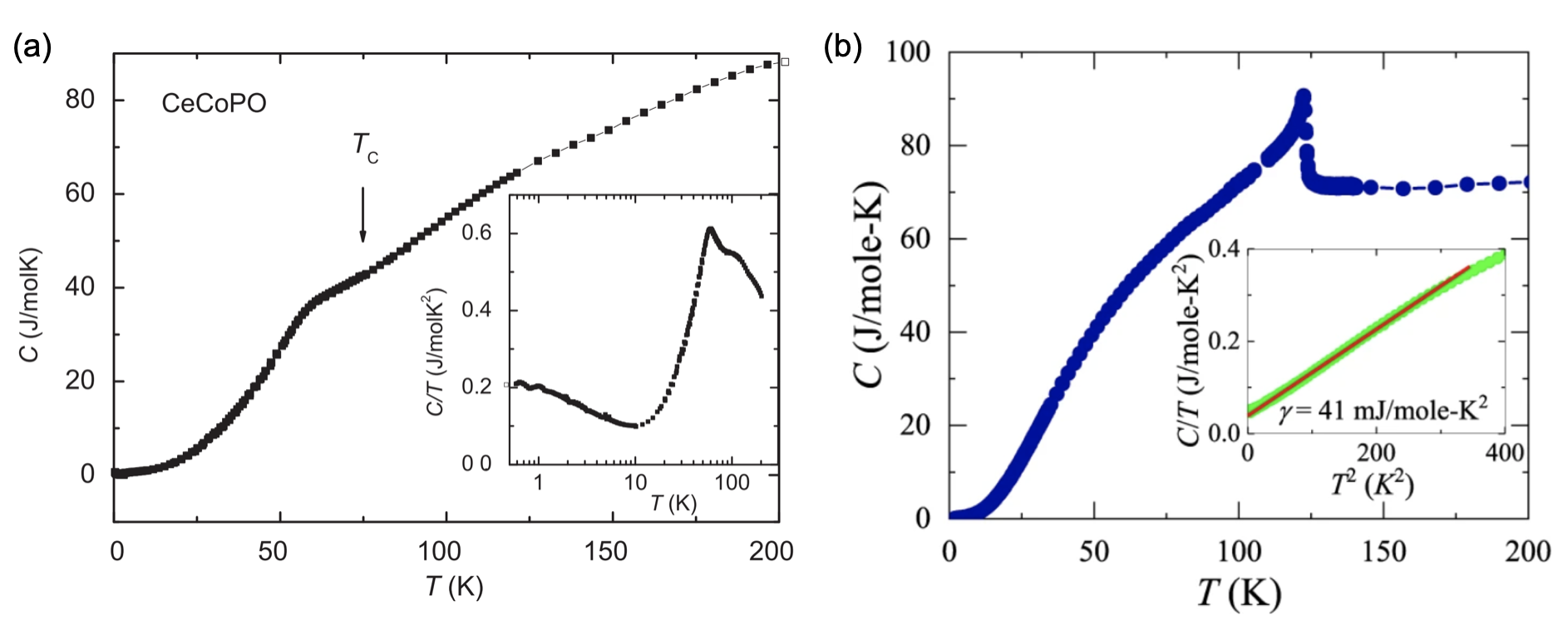}
    \caption{The specific heat of (a) $\rm CeCoPO$ copied from Ref.~\cite{CeCoPO} and (b) $\rm USbTe$ copied from Ref.~\cite{USbTe}.}
    \label{fig:specific_heat}
\end{figure}

We 
expect $\rm USbTe$ as a candidate of MWKSM with FM order.
It is likely that 
$\rm USbTe$
has two types of $f$ orbitals, with one orbital being
magnetically ordered while the other exhibiting the Kondo effect, in a way discussed before for similar U-based systems~\cite{Qiuyun-Chen-prl19}.
The ordered magnetic moment of $\rm USbTe$
is $\mu_s=0.4 \mu_B$~\cite{USbTe}.
The MSG is expected to be no.~129.417 ($P4/nm'm'$) supports WNLs as Fig.~\ref{fig:BZs} (b) shows. Assisted with the semimetallic resistivity (see Fig.~\ref{fig:resistivity} (c)) and enhanced $\gamma=41 {\rm mJ/K^2mol}$ (see Fig.~\ref{fig:specific_heat} (b)) comparing to its low carrier concentration $6\times 10^{20}~ {\rm cm}^{-3}$, $\rm USbTe$ is a good candidate of MWKSM. 
This material 
is expected to have symmetry enforced degeneracy at the Brillouin zone boundary and symmetry protected Weyl nodal lines as 
Fig.~\ref{fig:BZs} (b) shows.

We also consider $\rm CeMn_{2}Ge_{4}O_{12}$,
which has an AFM order on $\rm Mn$ atoms with MSG no.~125.367. In this MSG, the $\rm Ce$ atoms are always nonmagnetic and there are HWNLs; importantly, the symmetries allow the 3rd-order anomalous Hall response as the leading order. 
We propose to metallize this 
insulating material~\cite{xu2017magnetic} by chemical 
substitution or pressure to realize the topological state. 

\section{\label{app:materials} 
Additional materials as potential candidates}
\begin{table*}[ht]
    \centering
    \begin{tabular}{c|cc|c|c|c |c|cc|c}
         Materials & MSG \#&BNS  &Hall response & $T_{curie}$ & $T_N$ & $T_K$ &$\rho \rm(\mu\Omega cm)$ &RRR & $\gamma \rm(mJ /K^2mol)$\\
         \hline 
         $\rm CeMn_{2}Ge_{4}O_{12}$~\cite{xu2020high,xu2017magnetic} &125.367&$P4'/nbm'$  &3rd-order & &7.6K& &  & \\

         $\rm UNiGa$~\cite{UNiGa,UNiGamag} &189.224 &$P\bar{6}'2m'$  &3rd-order & &38K& &95 &3 &59 \\

         $\rm UNiAl$~\cite{UNiGa,UNiGamag} &189.224 &$P\bar{6}'2m'$  &3rd-order & &21K& &215 &1.1 &160 \\

         $\rm USb$~\cite{USb} &224.113 &$Pn\bar{3}m'$ &3rd-order & &213K & & & & \\

         $\rm UO_2$~\cite{authier2013international} &224.113 &$Pn\bar{3}m'$ &3rd-order & & 31K~\cite{UO2TN} & & & & \\
         \hline
          
         $\rm Ce_{4}Ge_{3}$~\cite{morozkin2009magnetic} &122.333 &$I\bar 42d$  &2nd-order & &6K,12K & & & &\\
          
         $\rm Ce_{4}Sb_{3}$~\cite{nirmala2009understanding} &122.336 &$I\bar 4'2d'$ &2nd-order &5K &$<$ 14K & & & & \\
          
         $\rm Ce_{3}NIn$~\cite{gabler2008magnetic} &117.305 &$P_C \bar 4 b2$  &2nd-order & &9K & & & &\\
         
         $\rm CeNiAsO$~\cite{wu2019incommensurate,luo2014heavy} &4.10 &$P_a2_1$&2nd-order & &8.7K &15K &$< 10$ & 100 &$203$\\
         
         $\rm CeSbTe$~\cite{schoop2018tunable} &4.10 &$P_a2_1$ &2nd-order & &2.7K & & & &$10$\\
         
         $\rm CeAuSb_2$~\cite{PhysRevB.106.224415,PhysRevB.93.195124} &39.201 &$A_bBm2$ &2nd-order & &6.5K &10K &$20 \sim 60$ &6 &$>50 $\\ 
         
         $\rm CeCuGa_3$~\cite{PhysRevB.104.174438} &44.230 &$Imm21'$&2nd-order &0.4K &1.9K &4-6K && &$>99$\\
         
         $\rm CeCu_2Ge_2$~\cite{knopp1989magnetic} &44.230 &$Imm21'$ &2nd-order & &4.1K &10K &$<120$~\cite{rauchschwalbe1985investigation} &1.1  &$1400$\\

         $\rm UCu_5$~\cite{UCu5} &161.72 &$R_I3c$ &2nd-order & & & &$\sim 160$~\cite{UCu5rho} &2.2 &\\

         $\rm UPtGe$~\cite{UPtGe} &44.234 &$I_amm2$ &2nd-order & &52K & &$\sim 200$~\cite{UPtGeres} &1.1  &\\

         \hline
         $\rm CeFeAsO$~\cite{zhao2008structural,dai2009f} &21.41 &$C22'2'$ &1st-order & &4K &10K ,130K&$7.5\times 10^4 $~\cite{chen2008superconductivity} & 0.8&$\sim 250 $~\cite{PhysRevB.81.134422}\\

         $\rm CeCoPO$~\cite{CeCoPO} &129.417&$P4/mm'm'$ &1st-order &75K & &40K & $500$ & 10&$200$ \\

         $\rm CeMn_2Ge_2$~\cite{welter1995neutron} &97.154 &$I42'2'$  &1st-order &300K &&& &\\

         $\rm Ce_2PdGe_3$~\cite{cedervall2016low} &6.20 &$Pm'$ &1st-order & &30K &&$200$ &1.8 &$50$~\cite{PhysRevB.91.035102}\\

         $\rm CeCoGe_3$~\cite{xu2020high} &107.231&$I4m'm'$  &1st-order & & & &  &&$111$~\cite{PhysRevB.47.11839}\\

         $\rm USbTe$~\cite{USbTe} &129.417&$P4/mm'm'$ &1st-order &125K & & &120 &1  &41 \\

         $\rm UBiTe$~\cite{UBiTe} &129.417&$P4/mm'm'$ &1st-order &105K & & &40 &3.8  &32 \\

         $\rm YbNiSn$~\cite{YbNiSn} &62.446 &$Pn'm'a$ &1st-order & &5.77K & &&&300~\cite{doi:10.1143/JPSJ.60.3145} \\

         $\rm CeCuGe$~\cite{CeAuCuGe} &11.54 &$P2_1'/m'$ &1st-order &10K & & &11 &1.6 &103~\cite{BMSondezi-Mhlungu_2010} \\

         \hline
         $\rm Ce_2RuZn_4$~\cite{Ce2RuZn4} &129.419&$P4'/nm'm'$ &no response & &2.5K & & & &30\\
         
        $\rm CeMnSbO$~\cite{CeMnSbO} &129.416&$P4'/n'm'm$ &no response & &240K & & & &\\

        $\rm CeMnSi$~\cite{CeMnSi} &59.407&$Pm'mn$ &no response & &130K &140 & 300 &8 &54 \\ 
    \end{tabular}
    \caption{Material Candidates of $\rm Ce$ and $\rm U$ based materials with (nonlinear) Hall responses. The transition temperatures of magnetic ordering due to $f$ electrons in $\rm Ce$ are listed. The resistivity $\rho$ at low temperature, residual resistivity ratio (RRR) and Sommerfeld constant $\gamma$ in heat capacity are also listed. 
    }
    \label{tab:candidate}
\end{table*}

\begin{table*}[ht]
    \centering
    \begin{tabular}{c|cc|c|c|c|c }
         Materials & MSG \#&BNS  &MPG &Hall response &Symmetry enforced &Accidental \\
         \hline 
         $\rm CeMn_{2}Ge_{4}O_{12}$~\cite{xu2020high,xu2017magnetic} &125.367&$P4'/nbm'$  &$4'/mmm'$ &3rd-order &HWNL  & \\

         $\rm UNiGa$~\cite{UNiGa,UNiGamag} &189.224 &$P\bar{6}'2m'$  &$\bar{6}'m'2$ &3rd-order &Weyl & \\

         $\rm UNiAl$~\cite{UNiGa,UNiGamag} &189.224 &$P\bar{6}'2m'$  &$\bar{6}'m'2$ &3rd-order &Weyl & \\

         $\rm USb$~\cite{USb} &224.113 &$Pn\bar{3}m'$  &$m\bar{3}m'$ &3rd-order &HWNL & \\

         $\rm UO_2$~\cite{authier2013international}  &224.113 &$Pn\bar{3}m'$  &$m\bar{3}m'$ &3rd-order &HWNL & \\
         \hline
          
         $\rm Ce_{4}Ge_{3}$~\cite{morozkin2009magnetic} &122.333 &$I\bar 42d$  &$\bar 42m$ &2nd-order &  &WNL\\
          
         $\rm Ce_{4}Sb_{3}$~\cite{nirmala2009understanding} &122.336 &$I\bar 4'2d'$  &$\bar 4'2m'$ &2nd-order &Dirac & \\
          
         $\rm Ce_{3}NIn$~\cite{gabler2008magnetic} &117.305 &$P_C \bar 4 b2$  & $\bar 4 2m1'$ &2nd-order &HWNL  &\\
         
         $\rm CeNiAsO$~\cite{wu2019incommensurate,luo2014heavy} &4.10 &$P_a2_1$ &$21'$ &2nd-order &Hourglass &\\
         
         $\rm CeSbTe$~\cite{schoop2018tunable} &4.10 &$P_a2_1$ &$21'$ &2nd-order &Hourglass &\\
         
         $\rm CeAuSb_2$~\cite{PhysRevB.106.224415,PhysRevB.93.195124} &39.201 &$A_bBm2$ &$mm21'$ &2nd-order &Weyl &\\ 
         
         $\rm CeCuGa_3$~\cite{PhysRevB.104.174438} &44.230 &$Imm21'$ &$mm21'$ &2nd-order &Weyl & \\
         
         $\rm CeCu_2Ge_2$~\cite{knopp1989magnetic} &44.230 &$Imm21'$ &$mm21'$ &2nd-order &Weyl &\\

         $\rm UCu_5$~\cite{UCu5} &161.72 &$R_I3c$ &$3m1'$ &2nd-order &Weyl & \\

         $\rm UPtGe$~\cite{UPtGe} &44.234 &$I_amm2$ &$mm21'$ &2nd-order &Weyl &\\

         \hline
         $\rm CeFeAsO$~\cite{zhao2008structural,dai2009f} &21.41 &$C22'2'$ &$22'2'$ &1st-order & &Weyl \\

         $\rm CeCoPO$~\cite{CeCoPO} &129.417&$P4/nm'm'$  &$4/mm'm'$ &1st-order & &WNL \\

         $\rm CeMn_2Ge_2$~\cite{welter1995neutron} &97.154 &$I42'2'$ &$42'2'$  &1st-order & &Weyl\\

         $\rm Ce_2PdGe_3$~\cite{cedervall2016low} &6.20 &$Pm'$ &$m'$ &1st-order & &Weyl\\

         $\rm CeCoGe_3$~\cite{xu2020high} &107.231&$I4m'm'$  &$4m'm'$ &1st-order & &Weyl \\

         $\rm USbTe$~\cite{USbTe} &129.417&$P4/nm'm'$  &$4/mm'm'$ &1st-order & &WNL \\

         $\rm UBiTe$~\cite{UBiTe} &129.417&$P4/nm'm'$  &$4/mm'm'$ &1st-order & &WNL \\

         $\rm YbNiSn$~\cite{YbNiSn} &62.446 &$Pn'm'a$  &$4'/mmm'$ &1st-order &Dirac, Weyl & \\

         $\rm CeCuGe$~\cite{CeAuCuGe} &11.54 &$P2_1'/m'$  &$4'/mmm'$ &1st-order &Weyl & \\

         \hline
         $\rm Ce_2RuZn_4$~\cite{Ce2RuZn4} &129.419&$P4'/nm'm'$  &$4'/mm'm'$ &no response & &Dirac\\
         
        $\rm CeMnSbO$~\cite{CeMnSbO} &129.416&$P4'/n'm'm$  &$4'/m'm'm$ &no response & &Dirac\\

        $\rm CeMnSi$~\cite{CeMnSi} &59.407&$Pm'mn$  &$m'mm$ &no response & &Dirac \\
         
    \end{tabular}
    \caption{Symmetry enforced crossings and symmetry allowed accidental crossings if there is no symmetry enforced crossings in the MSGs of the material candidates.
    }
    \label{tab:candidate_crossing}
\end{table*}

Beyond the primary materials candidates discussed earlier, here we discuss other potential candidates, even though the information about these additional candidate materials is more limited and, thus, the case for their candidacy for realizing MWKSMs is not as strong as for the ones discussed earlier (and in the main text). Still, we discuss them, as additional characterizations may strengthen the case.
Some interesting materials without anomalous Hall are also included.
In particular, we have searched MAGNDATA~\cite{gallego2016magndataI,gallego2016magndataII} and ICSD database.
The MSGs, order of anomalous Hall effect, transition temperature, Kondo temperature, resistivity at low temperature and Sommerfeld coefficient are summarized in Table~\ref{tab:candidate}.
The symmetry enforced crossings and symmetry allowed crossings are summarized in Table~\ref{tab:candidate_crossing}. 
The two tables are arranged in the order of the anomalous Hall effect.

We explain some other potential candidates listed in Table~\ref{tab:candidate}, that we cannot fully determine whether it is a MWKSM with current limited information.

$\rm UCu_5$~\cite{UCu5} and $\rm UPtGe$~\cite{UPtGe} are  semimetallic according to the resistivity measurement. However, the information about Kondo physics and correlation is not studied.

$\rm CeNiAsO$~\cite{wu2019incommensurate,luo2014heavy} has MSG is $P_a2_1$ (no.~4.10). The symmetry allowed leading order anomalous Hall response is 2nd-order. It has Neel temperature $T_N=8.7$K. The resistivity is metallic $\rho=10{\rm \mu\Omega/cm}$ and $\gamma=203{\rm mJ/K^2mol}$. 
The magnetic moments are at the $\rm Ce$ sites.

$\rm CeFeAsO$~\cite{zhao2008structural,dai2009f} is a well studied heavy fermion system with AFM order. Its MSG is $C22'2'$ (no.~21.41). The symmetry allows 1st-order anomalous Hall response. It has Neel temperature $T_N=130$K. The resistivity is semimetallic $\rho=2.8\times 10^3{\rm \mu\Omega/cm}$ and $\gamma\sim 250{\rm mJ/K^2mol}$. 
$\rm Ce$ can be magnetized in this MSG.
The MSG does not allow WNLs due to the absence of mirror symmetries. But accidental Weyl points can appear.

$\rm YbBiSn$~\cite{YbNiSn} has MSG $Pn'm'a$ (no.~62.446). This material has AFM order below $5.77$K with magnetic moment on $\rm Yb$ atoms. It has RRR $55-80$ and $\gamma=300{\rm mJ/K^2mol}$~\cite{doi:10.1143/JPSJ.60.3145}. The MSG enforces Dirac points and Weyl points on several high symmetry points.

$\rm CeCuGe$ is another possible candidate of MWKSM.
$\rm CeCuGe$~\cite{CeAuCuGe} has MSG $P2_1'/m'$ (no.~11.54). This material has FM order below $10$K with magnetic moment on $\rm Ce$ atoms. The resistivity has a semimetallic behaviour at low temperature $\rho \sim T^3$~\cite{CeAuCuGe}. $\gamma=109{\rm mJ/K^2mol}$~\cite{BMSondezi-Mhlungu_2010}. 
Besides, there are symmetry enforced Weyl points at high symmetry points in this MSG. 

$\rm UBiTe$ is very similar to $\rm USbTe$. They have the same MSG and similar resistivity and Sommerfeld coefficient. Therefore, it is likely also a MWKSM with FM order.

Multiple $f$-electrons sit at two different Wyckoff positions where one of them is ordered while the other exhibiting Kondo effect. The example is $\rm Ce_2RuZn_4$~\cite{Ce2RuZn4}. Ths MSG is $P4'/nm'm'$ (no.~129.419). The $\rm Ce$ at $2c$ Wyckoff position is magnetic while the $\rm Ce$ at $2a$ Wyckoff position is nonmagnetic.
It has Neel temperature $T_N=2.5K$. It is metallic with residual-resistance ratio RRR=2.5. It has $\gamma=30 {\rm mJ/K^2mol}$. 
There is no anomalous Hall effect and  no symmetry enforced crossings in this MSG.

$\rm CeMnSbO$~\cite{CeMnSbO} has MSG $P4'/n'm'm$ (no.~129.416) which contains $\cal PT$ symmetry. As a consequence, there is no anomalous Hall effect and all bands are Kramers degenerate. This material has metallic resistivity $\rho=30{\rm \mu\Omega cm}$ at $350$K and a semiconducting resistivity at low temperature. It has Neel temperature $T_N=240$K for $d$ electron AFM order. There is no data about Kondo effect~\cite{CeMnSbO}. The advantage of this material is that the MSG prevents $\rm Ce$ atom to be magnetized which makes the potential Kondo effect remain intact.

$\rm CeMnSi$~\cite{CeMnSi} has MSG $Pm'mn$ (no.~59.407) which contains $\cal PT$ symmetry. As a consequence, there is no anomalous Hall effect and all bands are Kramers degenerate. This material has metallic resistivity $\rho=300{\rm \mu\Omega cm}$ and large $\gamma=54 {\rm mJ/K^2mol}$. The Neel temperature is $T_N=242{\rm K}$. Therefore, there is coexistence of Kondo effect and AFM order~\cite{CeMnSi}. However, $\rm Ce$ can be magnetized in this MSG. The heavy fermion phase will be suppressed when the magnetic moment is large.

There are other potential materials candidates that can not be 
adequately assessed at the current stage due to the lack of information about their physical properties. Here we discuss some examples that are not listed in Table~\ref{tab:candidate}.
For example,
%
%
$\rm CeMnSb_2$~\cite{CeMnSb2} has an AFM order at 160K, which comes from $\rm Mn$ $d$ electrons, and Sommerfeld coefficient $\gamma=24.6 {\rm mJ/mol K^2}$~\cite{CeMnSb2}. However, since the experiment shows this material has a canted AFM order, there is not enough information to determine its MSG.
%

\end{document}